\documentclass[useAMS,usenatbib]{mn2e}
\usepackage{graphicx}
\usepackage{amssymb}
\usepackage{color}
\usepackage{hyperref}

\def\be{\begin{Equation}} 
\def\ee{\end{Equation}}

\def\msun{{\Msun}}

\def\HI{\hbox{H~$\scriptstyle\rm I$}} 
\def\HII{\hbox{H~$\scriptstyle\rm II$}} 
\def\HeI{\hbox{He~$\scriptstyle\rm I$}} 
\def\HeII{\hbox{He~$\scriptstyle\rm II$}} 
\def\HeIII{\hbox{He~$\scriptstyle\rm III$}}

\def\gsim{\lower.5ex\hbox{\gtsima}} 
\def\lsim{\lower.5ex\hbox{\ltsima}} \def\gtsima{$\; \buildrel > \over 
\sim \;$} \def\ltsima{$\; \buildrel < \over \sim \;$} \def\prosima{$\; 
\buildrel \propto \over \sim \;$} \def\gsim{\lower.5ex\hbox{\gtsima}} 
\def\lsim{\lower.5ex\hbox{\ltsima}} 
\def\simgt{\lower.5ex\hbox{\gtsima}} 
\def\simlt{\lower.5ex\hbox{\ltsima}} 
\def\simpr{\lower.5ex\hbox{\prosima}} \def\la{\lsim} \def\ga{\gsim} 
  
 \def\gtsima{$\; \buildrel > \over \sim \;$} 
\def\ltsima{$\; \buildrel < \over \sim \;$} 
\def\gsim{\lower.5ex\hbox{\gtsima}} 
\def\lsim{\lower.5ex\hbox{\ltsima}} 
\def\simgt{\lower.5ex\hbox{\gtsima}} 
\def\simlt{\lower.5ex\hbox{\ltsima}} 
\def\simpr{\lower.5ex\hbox{\prosima}} 
\def\la{\lsim} 
\def\ga{\gsim}

\def\msun{\,{\rm \Msun}}

\def\Msun{\rm M_\odot}

\def\avchi{$\langle \chi_\mathrm{HI} \rangle$}
 
\title[Simulating reionization]{The accuracy of semi-numerical reionization models in comparison with radiative transfer simulations}
\author[Hutter]{Anne Hutter$^{1,2}$\thanks{E-mail: ahutter@swin.edu.au}\\ 
$^{{1}}$ Swinburne University of Technology, Hawthorn, VIC 3122, Australia\\
$^{{2}}$ ARC Centre of Excellence for All Sky Astrophysics in 3 Dimensions (ASTRO 3D)}
\begin{document} 
 
\date{} 
 
 
\maketitle 
 
\label{firstpage} 
\begin{abstract} 
We have developed a modular semi-numerical code that computes the time and spatially dependent ionization of neutral hydrogen (\HI), neutral (\HeI) and singly ionized helium (\HeII) in the intergalactic medium (IGM). The model accounts for recombinations and provides different descriptions for the photoionization rate that are used to calculate the residual \HI\ fraction in ionized regions. 
We compare different semi-numerical reionization schemes to a radiative transfer (RT) simulation. We use the RT simulation as a benchmark, and find that the semi-numerical approaches produce similar \HII\ and \HeII\ morphologies and power spectra of the \HI\ 21cm signal throughout reionization. As we do not track partial ionization of \HeII, the extent of the double ionized helium (\HeIII) regions is consistently smaller. In contrast to previous comparison projects, the ionizing emissivity in our semi-numerical scheme is not adjusted to reproduce the redshift evolution of the RT simulation, but directly derived from the RT simulation spectra. Among schemes that identify the ionized regions by the ratio of the number of ionization and absorption events on different spatial smoothing scales, we find those that mark the entire sphere as ionized when the ionization criterion is fulfilled to result in significantly accelerated reionization compared to the RT simulation. Conversely, those that flag only the central cell as ionized yield very similar but slightly delayed redshift evolution of reionization, with up to 20\% ionizing photons lost. Despite the overall agreement with the RT simulation, our results suggests that constraining ionizing emissivity sensitive parameters from semi-numerical galaxy formation-reionization models are subject to photon nonconservation.
\end{abstract}


\section{Introduction}
\label{sec_intro}

The Epoch of Reionization (EoR) represents the last major phase transition in the Universe where it went from a neutral to an ionized state. It starts with the appearance of the first stars and galaxies, when high-energy photons permeate the intergalactic medium (IGM) and gradually ionize the hydrogen until the Universe is completely ionized by redshift $z\simeq6$ \citep{Fan2006}.

The exact growth of the ionized regions and the dependence of their sizes on time and location, i.e. its topology, remain outstanding questions. Based on simulations a picture has emerged, wherein, given the sources of ionizing radiation are located in high-density peaks, the ionization fronts originate in over-dense regions before percolating into under-densities \citep{Iliev2012, Battaglia2013, Bauer2015, Hutter2017, Finlator2009}. Though, recombinations can outweigh ionizations in the densest regions in the IGM, such that filaments and the outer parts of halos become self-shielded and remain at least partially neutral, creating so-called Lyman Limit systems \citep{Bolton_Haehnelt2007, Choudhury2009, Kakiichi2016}.

The reionization topology is strongly linked to the nature of the key ionizing sources. In the past decade observations and simulations increasingly supported a scenario in which star-forming galaxies are the main drivers of reionization \citep{Wyithe_Loeb2003, Choudhury_Ferrara2007, McQuinn2012}. Although observations of an unexpected high number of faint quasars by \citet{Giallongo2015} have questioned this picture, a number of studies have rebutted this finding. \citet{Parsa2017} could only confirm one of five $z>5$ X-ray detections and \citet{Weigel2015} yielded no quasar candidates at $z>5$ when observing the same field. 
\citet{Finlator2016} and \citet{Daloisio2017} found a quasar driven reionization scenario being disfavoured by the measured metal absorber abundances and Lyman-alpha forest measurements of the IGM temperature, respectively. Similarly, models of galaxy formation and reionization could not simultaneously fit the Thomson optical depth, the ionizing emissivity at $z\sim5$ and a high number of faint quasars at $z>5$ \citep{Qin2017,Hassan2017,Kulkarni2017}.

While the number of high-redshift sources are increasing \citep{Ouchi2017, Bouwens2017, Banados2016}, the contribution of star-forming galaxies to the ionizing budget remains still unclear due to the unconstrained faint end slope and cut-off of their luminosity function, as well as the unknown fraction of ionizing photons that escape from the galaxy into the IGM. In particular, the escape fraction of ionizing photons is a highly debated parameter that depends strongly on the inhomogeneous density structure of the interstellar medium in galaxies. Constraints from high-z galaxy observations yield average escape fractions of about 20\% \citep{Bouwens2016, Ishigaki2017}, and recent RT simulations find the escape fraction to be sensitive to supernovae explosions and radiative feedback \citep{Paardekooper2015, Kimm2014, Kimm2017}.

Since there is a lack of direct observations of the ionizing escape fraction from galaxies and reionization topology, these open questions have been addressed using a variety of simulations looking to find measurable signatures in the observables. 
The most accurate approach in describing the interplay between the ionization of the IGM and galaxy evolution are radiation hydrodynamical simulations \citep{Gnedin2014, Oshea2015, Ocvirk2016, Pawlik2017}. Yet these simulations are computationally very expensive, and in order to resolve the ionizing sources box sizes are comparatively small \citep[$\lsim90$cMpc in][]{Ocvirk2016}. Post-processing cosmological hydrodynamical or dark matter (DM) only simulations with RT \citep[e.g.][and references therein]{Trac2007, Ciardi2012, Hutter2014, Iliev2014} provides a cheaper alternative, as the RT calculations can be run with a coarser resolution, however, 3D RT remains computationally expensive. 

In contrast semi-numerical simulations that model the ionization of the IGM are computationally significantly cheaper. For example, \citet{Thomas2009} or \citet{Ghara2015a,Ghara2015b} yield ionization fields from mapping previously computed 1D ionization profiles of sources on a grid, while \citet{Furlanetto2004} identifies ionized regions by comparing the number of ionization and absorption events in different sized spheres. This filtering technique is independent of the number of ionizing sources and accounts for the ionizing radiation of adjacent sources. 

During the last few years multiple variations of the latter method have appeared, which either derive the number of ionizing photons from the collapsed fraction in each grid cell \citep{Furlanetto2004, Mesinger2007, Mesinger2011, Zahn2007, Zahn2011} or the halo/galaxy/stellar mass \citep{Santos2010, Kim2013, Mutch2016}, and use a tophat filter in real space \citep{Zahn2007,Santos2010, Mesinger2011, Mutch2016} or reciprocal space \citep{Zahn2011, Majumdar2014}, and flag the central cell \citep{Mesinger2011, Zahn2007, Zahn2011, Majumdar2014, Mutch2016} or the entire sphere \citep{Mesinger2007, Santos2010} as the ionization criterion is met at a particular smoothing scale.
All variations have been shown to produce power spectra of the \HI\ sensitive 21cm signal that are comparable to those of radiative transfer calculations, when their ionizing emissivity was tuned to reproduce the same volume or mass averaged neutral hydrogen fraction \citep{Zahn2011,Majumdar2014}. 

However, there are two issues with these existing semi-numerical methods. Firstly, either they assume \HeII\ (singly ionized helium) ionized regions to follow \HII\ regions \citep[e.g.][]{Mesinger2011,Santos2010}, or they focus solely on helium reionization \citep{Dixon2014}. Secondly, in comparison with radiative transfer runs, their ionizing emissivity or escape fraction of ionizing photons have always been adjusted to obtain the same average ionization fraction, such that deviations in the redshift evolution of reionization could not be revealed.

This paper addresses both issues. 
Firstly, we introduce our newly developed publicly available\footnote{The source code and instructions for compilation and execution can be found at \url{https://github.com/annehutter/grid-model}} MPI-parallelised semi-numerical reionization code that simultaneously simulates hydrogen and helium reionization and their interplay. It generates the number of ionizing photons according to the galaxies' spectra and follows the evolution of the fully \HII, \HeII, and \HeIII\ ionized regions. Particularly in the vicinity of sources with hard ionizing spectra, such as quasars or young, metal-poor stellar population, hydrogen and helium ionization fronts do not coincide and have an impact on the heating of the IGM as well as the recombination rates of the species. Following \HeII\ and \HeIII\ explicitly allows one to trace imprints of early quasars as well as modelling the hyperfine transition of the singly ionized isotope helium-3 ($^{3}$\HeII) \citep{McQuinn2009,Takeuchi2014}.
While helium reionization by quasars has been modelled by a combination of semi-analytic galaxy formation models and radiative transfer simulation \citep{LaPlante2016,LaPlante2017} and is certainly of interest around the peak of star formation, we focus in this paper on the epoch of hydrogen reionization. Since the contribution of quasars to reionization remains disputed, we consider a hydrogen reionization scenario driven by star-forming galaxies.

Secondly, similar to previous works, we evaluate the accuracy of two different semi-numerical reionization models by comparing their results to a radiative transfer simulation performed with \textsc{pcrash} \citep{Partl2011,Hutter2014}. However, we derive the ionizing emissivities in the semi-numerical schemes from the spectra of the sources that were used in the RT simulation, and do not adjust them to reproduce the redshift evolution of the RT simulation. Such a comparison gives insight into the precision with which e.g. semi-analytic galaxy formation models coupled to semi-numerical reionization schemes \citep[e.g.][]{Mutch2016} can constrain reionization sensitive parameters, such as the escape fraction of ionizing photons.

In Section \ref{sec_model} we describe the details of our modular semi-numerical reionization code; in particular the computation of recombinations, the photoionization rate and the residual \HI\ fraction in ionized regions, as well as the steps for computing the singly and double ionized helium regions in the IGM. Section \ref{sec_RT} describes the RT simulation that we use to compare our semi-numerical simulations to. In Section \ref{sec_comparison} we compare results from semi-numerical and the RT simulation. We conclude in Section \ref{sec_conclusions}.

\section{The semi-numerical scheme}
\label{sec_model}

In this Section we describe our semi-numerical model of the time and spatial evolution of the ionized regions in the IGM. Firstly, we describe the scheme to determine the ionized regions. Secondly, we present our model for the spatial dependent photoionization rate, the residual \HI\ fraction in ionized regions and inhomogeneous recombinations. Finally, we refine our scheme to identify the regions of singly and double ionized helium.

\subsection{Hydrogen ionized regions}

In order to calculate the distribution of the ionized regions in the IGM we follow the approach outlined in  \citet{Furlanetto2004}. This approach is based on the assumption that a region becomes fully ionized ($\chi_\mathrm{HII}=1$) as the number of ionizing photons emitted exceeds the number of absorptions in this region, otherwise it remains completely neutral ($\chi_\mathrm{HII}=0$). It automatically accounts for the radiation from multiple neighbouring ionizing sources when the chosen region is large enough. 

Following \citet{Furlanetto2004} we apply a filtering method to our density and ionizing emissivity grids: We convolve the ionizing emissivity and gas density fields with real-space top-hat filters of subsequently decreasing size, which corresponds to averaging those quantities over a spherical region with radius $R$. If the number of ionizing photons exceeds the number absorbed in this smoothed region, the central cell is marked as ionized \citep[as in e.g.][]{Zahn2007, Zahn2011,Mesinger2011, Majumdar2014, Mutch2016}. 
An alternative approach is to mark the entire smoothing sphere as ionized \citep{Mesinger2007,Santos2010}, which we refer to as the ``ionized sphere'' flag
in the following\footnote{Our code has the option of choosing between either approaches, i.e. marking the central cell (default) or the entire smoothing sphere (``ionized sphere'') as ionized.}.
The radius of the smoothing sphere is subsequently decreased, and cells with a balance of ionization and absorption events larger than unity become ionized. This procedure - starting at a maximum value $R$ and subsequently decreasing it to cell size - accounts in each cell for the ionizing emissivity originating from neighbouring sources. At the smallest smoothing scale, i.e. cell size, we allow cells to be partially ionized with their ionization fraction given by the ratio between ionization and absorption events, similar to \citet{Majumdar2014, Kim2013} and \citet{Choudhury2009}.

The criterion whether a cell is ionized or not depends whether the cumulative number of ionizing photons,
\begin{eqnarray}
 N_\mathrm{ion}(z) &=& \sum_{i=1}^{N_\mathrm{s,cell}}\ \int_z^{z_\mathrm{form,i}} \dot{N}_\mathrm{ion,i}(z') \frac{\mathrm{d}t}{\mathrm{d}z'}\ \mathrm{d}z' \\
 &=& \sum_{i=1}^{N_\mathrm{s,cell}} \int_z^{z_\mathrm{form,i}} \int_{\nu_\mathrm{HI}}^{\infty} \frac{L_{\nu\mathrm{,i}}(z')}{h\nu}\ \frac{\mathrm{d}t}{\mathrm{d}z'}\ \mathrm{d}\nu\ \mathrm{d}z', \nonumber
 \label{eq_subsec_esf_nion}
\end{eqnarray}
exceeds the number of absorptions, including those from recombination events,
\begin{eqnarray}
 N_\mathrm{ion}(z) &\geq& N_\mathrm{abs}(z) = \langle n_\mathrm{H,0}\rangle_R\ V_\mathrm{cell}\ \left[1 + \bar{N}_\mathrm{rec}(z)\right].
 \label{eq_subsec_esf_ioncriterion}
\end{eqnarray}
Here $L_{\nu\mathrm{,i}}$ represents the spectrum of source $i$ of $N_\mathrm{s,cell}$ sources in the cell and $z_\mathrm{form,i}$ the redshift of its onset. $\langle n_\mathrm{H,0}\rangle_R$ is the hydrogen number density today smoothed over a sphere with radius $R$, and $V_\mathrm{cell}$ the comoving volume of the grid cell.

\subsection{Hydrogen photoionization rate}
\label{sec_hdro_photion}

We aim to derive a description for a spatially dependent \HI\ photoionization rate, which solely depends on the distribution of ionizing sources and the respective sizes of ionized regions. This description will be essential to compute the \HI\ residual fraction in ionized ($\chi_\mathrm{HII}>0.99$) regions below, and to derive the radiative feedback on star formation in galaxies when combining this reionization model with semi-analytic galaxy formation models. Before describing the details of our photoionization rate model, we clarify its definition and how different models can be deduced from it.
The photoionization rate represents the number of ionization events per unit time for the respective species, and is given by the incident rate of ionizing photons ($\nu>\nu_s$) per unit area, $F_\nu/h\nu$, and the species ionization cross section, $\sigma_\mathrm{s}(\nu)$.
\begin{eqnarray}
 \Gamma_\mathrm{s}&=& \int_{\nu_\mathrm{s}}^{\infty} \mathrm{d}\nu\ \sigma_\mathrm{s}(\nu)\ \frac{F_\nu}{h\nu} =\int_{\nu_\mathrm{s}}^{\infty} \mathrm{d}\nu\ \sigma_\mathrm{s}(\nu)\ \epsilon_\nu\ \lambda_\mathrm{mfp}(\nu).
 \label{eq_subsec_photion_Gamma}
\end{eqnarray}
The incident rate per unit area can also be written in terms of the specific emissivity $\epsilon_\nu$ and the mean free path $\lambda_\mathrm{mfp}(\nu)$. 
Computing the spatially and time dependent photoionization rate is expensive, since the ionizing flux (or mean free path) depends on the neutral gas density in all cells between all ionizing sources and the cell considered. Thus we pursue two different approximations: we either deduce the spatially dependent mean free path from the extent of the ionized regions (approach 1, see e.g. \citet{Sobacchi2014, Mutch2016}), or generalise the decline of the ionizing flux with distance from the source (approach 2, see e.g. \citet{Mesinger2015}). Both approaches are reflected in the right and left hand side of Equation \ref{eq_subsec_photion_Gamma}, respectively.
Throughout this Section, the ionizing cross section for \HI\ is given by $\sigma_\mathrm{HI}(\nu)=\sigma_\mathrm{HI,0} (\nu/\nu_\mathrm{HI})^{-\beta}$ and the ionizing emissivity by $\epsilon_\nu = \epsilon_\mathrm{HI} (\nu/\nu_\mathrm{HI})^{-\alpha-1}$, with $\nu_\mathrm{HI}=13.6$eV, $\beta=3$ and $\alpha$ being the spectral index of the ionizing sources. The mean free path $\lambda_\mathrm{mfp}$ is assumed to be constant.

\subsubsection{Mean free path based (first) approach}
\label{sec_model_par_photHI_model1}

In our first approach, we perform the frequency integration of the right hand side of Equation \ref{eq_subsec_photion_Gamma}. We find for the \HI\ photoionization rate,
\begin{eqnarray}
 \Gamma_\mathrm{HI}(\mathbf{x},z)&\simeq& \lambda_\mathrm{mfp}(\mathbf{x})\ \sigma_\mathrm{HI,0}\ \frac{\alpha}{\alpha+\beta}\ \frac{\dot{N}_\mathrm{ion,cell}(\mathbf{x},z)}{V_\mathrm{cell}(z)},
\end{eqnarray}
where $\dot{N}_\mathrm{ion,cell}(z)$ is the rate of incident ionizing photons in a chosen cell at redshift $z$, $V_\mathrm{cell}(z)$ is the corresponding physical volume of that cell, and $\dot{N}_\mathrm{ion,cell}/V_\mathrm{cell}$ depicts the total physical ionizing emissivity. 
The filtering method allows us to determine the physical mean free path $\lambda_\mathrm{mfp}$ for each cell as the largest filtering scale $R$ at which the cell becomes ionized. Correspondingly, $\dot{N}_\mathrm{ion,cell}$ represents the maximum incident rate of ionizing photons in the cell $max(\dot{N}_\mathrm{ion,cell}(r\leq R))$, when the ionizing emissivity field is smoothed over scales $r<R$. 
The drawback of this approach lies in its dependency on the accurate modelling of the mean free path. For example, in this model, initially small ionized regions that have recently merged with larger ionized region have smaller mean free path due to their smaller filtering scales; hence, there is a jump in the mean free path at the intersection of the initially small and large ionized regions. This may lead to a significant overestimate of the neutral hydrogen gas density in the immediate vicinity of sources in small ionized regions.

\subsubsection{Flux based (second) approach}
\label{sec_model_par_photHI_model2}

In the second approach we use the description of the ionizing flux as a function of the distance from a source. The ionizing flux decreases in proportion to $\propto e^{-\tau} r^{-2}$, with $r$ being the distance from the source and $\tau$ the optical depth of the ionizing radiation. In each cell the \HI\ photoionization rate is then the sum over the radiation fields of all ionizing sources ($N_s$),
\begin{eqnarray}
 \Gamma_\mathrm{HI}(\mathbf{x}) &=& \langle\sigma_\mathrm{HI}\rangle_{\nu}\ \frac{\alpha}{\alpha+\beta}\ \sum_{i=1}^{N_s} \frac{ \dot{N}_{\mathrm{ion,}i}}{4\pi|\mathbf{x}-\mathbf{x}_i|^2}\ e^{-\frac{|\mathbf{x}-\mathbf{x}_i|}{\lambda_\mathrm{mfp}}}\nonumber\\
 &&\times (1-e^{-1})^{-1} f^{-3},
\end{eqnarray}
where $\langle\sigma_\mathrm{HI}\rangle_{\nu}$ is the spectrum averaged \HI\ photoionization cross section, $\dot{N}_\mathrm{ion,i}$ the rate at which source $i$, located at $\mathbf{x_i}$, emits ionizing photons, and $f$ is a normalisation factor.
The optical depth $\tau$ is determined by the mean free path $\lambda_\mathrm{mfp}$ during reionization. $\lambda_\mathrm{mfp}$ is either given by the size of the ionized regions during reionization, $\lambda_\mathrm{mfp}^\mathrm{ion}$, or the distance between self-shielding regions in the final stages of reionization and post-reionization, $\lambda_\mathrm{mfp}^\mathrm{ss}$. 
In our model we assume $\lambda_\mathrm{mfp}=min(\lambda_\mathrm{mfp}^\mathrm{ion},\lambda_\mathrm{mfp}^\mathrm{ss})$, whereas 
$\lambda_\mathrm{mfp}^\mathrm{ion}$ is determined as the largest filtering scale, $R$, at which the cell becomes ionized. $\lambda_\mathrm{mfp}^\mathrm{ss}$ is computed as in \citet{Miralda2000},
\begin{eqnarray}
 \lambda_\mathrm{mfp}^\mathrm{ss} &=& \lambda_0 \left[ 1-F_\mathrm{V}(\Delta<\Delta_\mathrm{ss}) \right]^{-2/3},
\end{eqnarray}
with $\lambda_0=60/H(z)$~Mpc, $\Delta_\mathrm{ss}$ being the density below which the gas is ionized, and $F_\mathrm{V}$ the corresponding volume fraction that remains neutral. 
In reality, $\lambda_\mathrm{mfp}$ is light-of-sight dependent, and varies for each source according to the size of its surrounding ionized region and its degree of ionization. However, such a calculation can only be dealt with in radiative transfer simulations and is computationally expensive. Even when assuming a single $\lambda_\mathrm{mfp}$ value for each source the computation time scales as $\mathcal{O}(N_\mathrm{cell} N_s)$. 
Thus, we assume a global value for $\lambda_\mathrm{mfp}$ which is the median of the mean free paths in all ionized cells.\footnote{Choosing a single value for the mean free path allows us  to optimise our code using fast Fourier transformations} The relation between the mean and the median depends on the ionized bubble size distribution. To retain short computation times we assume the median to be $\sim70$\% of the mean, which corresponds to a log-normal distribution \citep{McQuinn2007, Meerburg2013} with $\sigma_{\ln R}\simeq0.9$.\footnote{The relation between the mean and the median depends on the ionized bubble size distribution and can be chosen in the code.}

We normalise the photoionization rate by adapting the factor $f$ such that the photoionization rate within a sphere of radius $\tilde{\lambda}_\mathrm{mfp}$ is equal to the value of our first approach. Further details are described in Appendix \ref{sec_app2}. 
In the remainder of the paper we will use the flux-based approach for the reasons described above. However, in Appendix \ref{sec_app_honly_comp_photionmodels} we compare the results of both approaches on the \HI\ residual fields within the ionized regions.

\subsection{Self-shielding of hydrogen}
\label{sec_selfshielding}

From the photoionization rate we can derive the residual neutral hydrogen fraction within the ionized regions. Since the typical cell size in a cosmological simulation is too coarse to resolve partially neutral self-shielded systems during reionization, we use a subgrid model for the gas density distribution within each cell. Furthermore, we base our calculations of the residual neutral hydrogen fraction on the modified photoionization rate in \citet{Rahmati2013}, which has been derived from well-resolved radiative transfer simulations.

Firstly, we estimate the density above which the hydrogen gas becomes self-shielded. For that purpose we consider the neutral hydrogen column density to be given by the neutral hydrogen on a Jeans-length scale according to \citet{Schaye2001}
\begin{eqnarray}
 N_{\mathrm{HI}}&=& \chi_{\mathrm{HI}}\ n_\mathrm{H}\ L_\mathrm{J} \nonumber \\
 &\simeq& \chi_{\mathrm{HI}} \left( \frac{\pi \gamma k_\mathrm{B}}{\mu m_\mathrm{H}^2 G} \right)^{1/2} \left( 1-Y \right)^{1/2} f_\mathrm{g}^{1/2} n_\mathrm{H}^{1/2} T^{1/2}.
 \label{eq_subsec_photion_NHI}
\end{eqnarray}
Since we consider gas within ionized regions, we can assume that the gas is mostly ionized and thus $n_\mathrm{HII}\simeq n_\mathrm{H}$ holds. The neutral hydrogen fraction is then determined by the recombination rate $\beta_\mathrm{HII}$ and the photoionization rate $\Gamma$,  $\chi_\mathrm{HI} = n_\mathrm{H} \beta_\mathrm{HII} \Gamma^{-1} (1-3Y/4) / (1-Y)$. The factor $(1-3Y/4) / (1-Y)$ accounts for the increased number of electrons available for recombinations due to singly ionized helium, with $Y$ being the helium mass fraction of the pristine gas.
The hydrogen gas density $n_\mathrm{H}$ is given by the local over-density, $\Delta$, and the corresponding mean hydrogen density in the Universe at the corresponding redshift, $\bar{n}_\mathrm{H}=3 H_0^2/(8\pi G m_\mathrm{H})\ \Omega_b (1-Y) (1+z)^3 \Delta$. $f_g$ is the baryon fraction of the matter density $\Omega_b/\Omega_m$. Inserting these relations into Equation \ref{eq_subsec_photion_NHI} yields
\begin{eqnarray}
 N_\mathrm{HI} &=& \left( \frac{3H_o^2\Omega_b (1+z)^3 \Delta}{8\pi G} \right)^{3/2} \left(\frac{\pi \gamma k_\mathrm{B}}{\mu m_\mathrm{H}^5 G} \right)^{1/2} \nonumber\\ 
 && (1-3Y/4) (1-Y) f_\mathrm{g}^{1/2} T^{1/2}\ \beta_\mathrm{HII}\ \Gamma^{-1}.
\end{eqnarray}
If the distance between the hydrogen atoms is smaller than their ionizing mean free path ($N_\mathrm{HI} \gtrsim \sigma_\mathrm{HI}^{-1}$), the ionizing radiation can not ionize each hydrogen atom, and the respective region becomes self-shielded; a region becomes self-shielded when its density $\Delta = n_\mathrm{H} / \bar{n}_\mathrm{H}$ exceeds
\begin{eqnarray}
 \Delta_\mathrm{ss}&=& \left( \frac{\mu m_\mathrm{H}^2 G}{\pi \gamma k_\mathrm{B} \sigma_\mathrm{HI}^2} \right)^{1/3} \left( 1-3Y/4 \right)^{-2/3} \left( 1-Y \right)^{1/3}\nonumber\\
 && T^{-1/3} f_\mathrm{g}^{-1/3} \beta_\mathrm{HII}^{-2/3} \Gamma^{2/3} \bar{n}_\mathrm{H}^{-1},
\end{eqnarray}
whereas $\sigma_\mathrm{HI}=\langle\sigma_\mathrm{HI}\rangle_{\nu}$ represents the mean photoionization cross section for all hydrogen ionizing photons at different energies and thus is dependent on the spectrum of the ionizing radiation.

Secondly, we consider the ``initial'' photoionization, $\Gamma$, as a background photoionization rate and compute the modified photoionization rate, $\Gamma_\mathrm{mod}$, in each cell. This modified photoionization rate accounts for the reduced ionization of the cell due to self-shielding. We use the relation found in \citet{Rahmati2013},
\begin{eqnarray}
 \frac{\Gamma_\mathrm{mod}}{\Gamma} &=& 0.98 \left[ 1+\left(\frac{\Delta}{\Delta_\mathrm{ss}}\right)^{1.64} \right]^{-2.28}+ 0.02 \left[ 1+\frac{\Delta}{\Delta_\mathrm{ss}} \right]^{-0.84},
\end{eqnarray}
and compute the neutral hydrogen ionization fraction assuming an equilibrium between the ionization and recombination events ($\mathrm{d}n_\mathrm{HI}/\mathrm{d}t=0$). 
\begin{eqnarray}
 \chi_\mathrm{HI} &=& 1+\frac{x(\Delta)}{2} \left[ \sqrt{1+\frac{4}{x(\Delta)}} - 1 \right], \\
 \mathrm{with}\ \ x(\Delta) &=&\frac{\Gamma_\mathrm{mod}}{\beta_\mathrm{HII} \bar{n}_\mathrm{H} \Delta}\ \frac{1-Y}{1-3Y/4}.
\end{eqnarray}
Our model of a spatially dependent photoionization rate and self-shielded regions allows us to derive the residual \HI\ fraction in regions flagged as ionized.

\subsection{Hydrogen recombinations}
\label{sec_model_recombinations}

The propagation of the ionized regions depends sensitively on the underlying gas density. In particular in high-density regions, recombinations can become significant and delay the ionization of these regions. 
Thus we derive the number of recombinations in each cell from the rate equation for ionized hydrogen,
\begin{eqnarray}
 \frac{\mathrm{d}n_\mathrm{HII}}{\mathrm{d}t}&=& \Gamma_\mathrm{HI}\ n_\mathrm{HI} - \beta_\mathrm{HII}\ n_\mathrm{HII}^2\ \chi_\mathrm{e}.
\end{eqnarray}
For a nearly completely ionized cell ($n_\mathrm{HI}\simeq0$), this equation simplifies and is analytically solvable. The difference between the initial and final \HII\ number densities at times $t_0$ and $t$ allows us to calculate the number of recombinations as
\begin{eqnarray}
 N_\mathrm{rec,HII}(t,t_0) &=& \left[1 + \frac{1}{n_\mathrm{HII}(t_0)\ \chi_\mathrm{e}\ \beta_\mathrm{HII}\ (t-t_0)}\right]^{-1},
 \label{eq_subsec_Nrec_HII}
\end{eqnarray}
where $\chi_\mathrm{e} = n_\mathrm{e}/n_\mathrm{HII}$ represents the fraction of electrons coming from hydrogen ionization. Including helium, $\chi_e=(1-3Y/4)/(1-Y)$ yields $1.08$ during reionization for $Y=0.25$.
In our model the number of recombinations are determined for each cell individually, where $t_0$ is the time or redshift when the cell was ionized and $t>t_0$ represents the time at which the number of recombinations are estimated. Since our description is independent from the photoionization rate, we can perform this calculation at each time before the ionized regions are identified with the filtering method.

Our model is very similar to \citet{Sobacchi2014}, though their derivation of the number of recombinations exploits the model described in \citet{Miralda2000} with the cell's density determining the underlying density distribution. 
Conversely, the model presented in \citet{Choudhury2009} is quite different. While in our model the number of recombinations in each cell depends on the time of ionization and its density, \citet{Choudhury2009} assumes a global value for the time of ionization. Furthermore, despite using the same condition for a region to be self-shielded, our scheme derives a residual \HI\ fraction, while \citet{Choudhury2009} only distinguishes between neutral or completely ionized cells.

\subsection{Including helium ionizations}
\label{sec_helium_ionization}

Our model can also derive the distribution of singly and double ionized helium regions. 
In analogy to hydrogen we compute the \HeII\ and \HeIII\ regions by comparing the number of helium ionizing photons and the number of absorptions (number of helium atoms plus their number of recombinations) for a range of different smoothing scales $R$.
The number of helium ionizing photons with $X=$\HeI, \HeII\ is given by 
\begin{eqnarray}
 N_\mathrm{X,ion}(z) &=& \sum_{i=1}^{N_\mathrm{s,cell}}\ \int_z^{z_\mathrm{form,i}} \dot{N}_\mathrm{X,ion,i}(z') \frac{\mathrm{d}t}{\mathrm{d}z'}\ \mathrm{d}z', 
 \label{eq_subsec_helium_nion}
\end{eqnarray}
and a cell becomes ionized when
\begin{eqnarray}
 N_\mathrm{X,ion}(z) &\geq& N_\mathrm{X,abs}(z) = \langle n_\mathrm{X,0}\rangle_R\ V_\mathrm{cell}\ \left[1 + \bar{N}_\mathrm{X,rec}(z)\right].
 \label{eq_subsec_helium_ioncriterion}
\end{eqnarray}
We assume that regions can be either neutral ($\chi_\mathrm{HeI}=0$), singly ionized ($\chi_\mathrm{HeII}=0$) or double ionized ($\chi_\mathrm{HeIII}=0$).

The number of helium ionizing photons depends strongly on the spectra of the sources. As spectra become harder, not only do the numbers of helium ionizing photons increase, but also a larger fraction of ionizing photons are absorbed by helium than by hydrogen. In the following paragraphs we describe our computation of the number of \HeI\ and \HeII\ ionizing photons.

\subsubsection{Deriving $\dot{N}_\mathrm{HeI,ion}$}
\label{sec_helium_nion}

While hydrogen recombination events result in photons with energies below $13.6$~eV that can neither ionize hydrogen nor helium, recombinations of singly ionized helium atoms produce photons that can ionize hydrogen and helium. For this reason the on the spot approximation (case B recombination) can not be used.

The \HeII\ recombination rate is given by
\begin{eqnarray}
 \beta_\mathrm{HeII} &=& y\ \beta_\mathrm{1,HeII} + \beta_\mathrm{B,HeII},
\end{eqnarray}
where $\beta_\mathrm{1,HeII}$ refers to all recombination events that end in the ground state of helium ($n=1$), and $\beta_\mathrm{B,HeII}$ accounts for the remaining transitions to excited states ($n>1$). $y$ is the fraction of photons produced by recombinations to the $n=1$ state that ionize hydrogen, while the remaining fraction $(1-y)$ ionizes helium. Due to this fact $\beta_\mathrm{HeII}$ is lower than the case A recombination rate ($\beta_\mathrm{A,HeII}=\beta_\mathrm{1,HeII} + \beta_\mathrm{B,HeII}$). The fraction $y$ depicts the ionization probability of hydrogen at $\nu_\mathrm{HeI}$ and depends on the overall helium mass fraction of the gas $Y$:
\begin{eqnarray}
 y(Y)&=& \frac{n_\mathrm{H}\ \sigma_\mathrm{HI}(\nu_\mathrm{HeI})}{n_\mathrm{H}\ \sigma_\mathrm{HI}(\nu_\mathrm{HeI}) + n_\mathrm{He}\ \sigma_\mathrm{HeI}(\nu_\mathrm{HeI})} \nonumber \\
 &=& \left[1+\frac{Y}{4(1-Y)}\frac{\sigma_\mathrm{HeI}(\nu_\mathrm{HeI})}{\sigma_\mathrm{HI}(\nu_\mathrm{HeI})}\right]^{-1}.
\end{eqnarray}
Hence, the number of helium and hydrogen ionizations are $Q_1(1-y)/(n_\mathrm{e} \beta_\mathrm{HeII})$ and $Q_0/(n_\mathrm{e} \beta_\mathrm{B,HII})$, respectively. $Q_0$ and $Q_1$ represent the number of photons with $\nu>\nu_\mathrm{Hi}$ and $\nu>\nu_\mathrm{HeI}$. For a spectrum $L_{\nu} \propto \nu^{-\alpha}$, the fraction $Q_1/Q_0$ yields $(\nu_\mathrm{HeI}/\nu_\mathrm{HI})^{-\alpha}$, and the relation between the number of helium and hydrogen ionizations or ionization rates is given by
\begin{eqnarray}
 \frac{\dot{N}_\mathrm{HeI,ion}}{\dot{N}_\mathrm{HI,ion}} &=& (1-y(Y)) \frac{\beta_\mathrm{B,HII}}{\beta_\mathrm{B,HeII}+y(Y)\ \beta_\mathrm{1,HeII}} \left(\frac{\nu_\mathrm{HeI}}{\nu_\mathrm{HI}}\right)^{-\alpha}. \nonumber\\
 \label{eq_subsec_helium_h_to_he_ratio}
\end{eqnarray}

\subsubsection{Deriving $\dot{N}_\mathrm{HeII,ion}$}
\label{sec_helium_doublenion}

Since the ionization cross sections for \HI\ and \HeI\ are a magnitude smaller than for \HeII\ at photon energies of $E>h\nu_\mathrm{HeII}=54.4$~eV, we assume the number of \HeII\ ionizing photons from a source with specific luminosity $L_\nu$ to be given by
\begin{eqnarray}
\dot{N}_\mathrm{HeII,ion} &=& \int_{\nu_\mathrm{HeII}}^{\infty} \frac{L_{\nu}(z')}{h\nu}\ \mathrm{d}\nu.
\end{eqnarray}
For starburst galaxies the number of \HeII\ ionizing photons can be about a ten thousandth or even less than the number of the \HI\ ionizing photons, while for quasars the numbers can be nearly of comparable size.

\subsubsection{Helium recombinations}
\label{sec_helium_reccombinations}

We compute the recombinations of \HeII\ and \HeIII\ analogous to \HII. The rate equations
\begin{eqnarray}
  \frac{\mathrm{d}n_\mathrm{HeII}}{\mathrm{d}t}&=& \Gamma_\mathrm{HeI}\ n_\mathrm{HeI} - \beta_\mathrm{HeII}\ n_\mathrm{HeII}^2\ \chi_\mathrm{HeII,e} + \beta_\mathrm{HeIII}\ n_\mathrm{HeIII}\ n_\mathrm{e} \nonumber\\
  &&
 \end{eqnarray}
 and
 \begin{eqnarray}
  \frac{\mathrm{d}n_\mathrm{HeIII}}{\mathrm{d}t}&=& \Gamma_\mathrm{HeII}\ n_\mathrm{HeII} - \beta_\mathrm{HeIII}\ n_\mathrm{HeIII}^2\ \chi_\mathrm{HeIII,e}
\end{eqnarray}
become analytically solvable for $n_\mathrm{HeI}\simeq0$, and either $n_\mathrm{HeIII}\simeq0$ or $n_\mathrm{HeII}\simeq0$. Thus, we find the number of recombinations between the initial and final \HeII\  and \HeIII\ number densities at times $t_0$ and $t$ to be 
\begin{eqnarray}
  N_\mathrm{rec,HeII}(t,t_0) &=& \left[1 + \frac{1}{n_\mathrm{HeII}(t_0)\ \chi_\mathrm{HeII,e}\ \beta_\mathrm{HeII}\ (t-t_0)}\right]^{-1}\nonumber\\
\end{eqnarray}
and
\begin{eqnarray}
  N_\mathrm{rec,HeIII}(t,t_0) &=& \left[1 + \frac{1}{n_\mathrm{HeIII}(t_0)\ \chi_\mathrm{HeIII,e}\ \beta_\mathrm{HeIII}\ (t-t_0)}\right]^{-1}. \nonumber\\
\end{eqnarray}
$\chi_\mathrm{e,HeII}$ and $\chi_\mathrm{e,HeIII}$ depend on the ionization state of hydrogen and the helium abundance in the Universe, and are given by $4(Y^{-1}-1)\chi_\mathrm{HII}+1$ and $4(Y^{-1}-1)\chi_\mathrm{HII}+2$, respectively.

The inclusion of helium also affects the number of \HII\ recombinations. Here, $\chi_\mathrm{e}$ in Equation \ref{eq_subsec_Nrec_HII} can adopt three values corresponding whether helium is not ionized ($\chi_\mathrm{HeII}=\chi_\mathrm{HeII}=0$), singly ionized ($\chi_\mathrm{HeII}=1$, $\chi_\mathrm{HeIII}=0$) or double ionized ($\chi_\mathrm{HeII}=0$, $\chi_\mathrm{HeIII}=1$):
\begin{eqnarray}
 \chi_\mathrm{e}&=& 1\ +\ \frac{Y}{4(1-Y)} \left[ \frac{\chi_\mathrm{HeII}}{\chi_\mathrm{HII}}\ +\ 2\ \frac{\chi_\mathrm{HeIII}}{\chi_\mathrm{HII}} \right].
\end{eqnarray}

\section{Radiative transfer simulation}
\label{sec_RT}

In this Section we describe the radiative transfer simulation of cosmic reionization, to which we compare our semi-numerical reionization schemes in Section \ref{sec_comparison}. We briefly summarise the simulation and refer the interested reader to \citet{Hutter2015} for further details.
In order to simulate reionization, a hydrodynamical simulation was post-processed with the MPI-parallelised 3D Monte-Carlo radiative transfer code \textsc{pcrash} \citep{Ciardi2001, Maselli2003, Maselli2009, Partl2011, Hutter2014}. \textsc{pcrash} is capable of treating multiple source spectra, a spatially dependent clumping factor, and yields both the evolving IGM (hydrogen and helium) ionization and temperature fields. For the respective simulation, \textsc{pcrash} computes the spatial and time evolution of hydrogen and helium.

The underlying hydrodynamical simulation was carried out with the TreePM-SPH code 
\textsc{gadget}-2 
\citep{Springel2005} and includes star formation, radiative and Compton cooling, a UV background switched on at $z=6$, supernova feedback, and stellar winds.
The adopted cosmological dark energy, matter and baryonic density parameters are $\Omega_\Lambda=0.73$, $\Omega_m=0.23$, $\Omega_b=0.047$, and a Hubble constant with a value of $H_0=100h=70$km~s$^{-1}$Mpc$^{-1}$. The simulation box has an edge length of $80h^{-1}$ comoving Mpc, contains $1024^3$ dark matter (DM) particles, and initially the same number of gas particles. The DM and gas particle masses are $3.6\times10^7h^{-1}\msun$ and $6.3\times10^6h^{-1}\msun$, respectively. Halos are identified as gravitationally bound groups of at least 20 particles using the Amiga Halo Finder \citep{Knollmann2009}. 

The snapshots of the hydrodynamical simulation between $z\simeq13$, when the first galaxies are identified, to $z\simeq6$ are consecutively post-processed with \textsc{pcrash}. We use only ``well-resolved'' galaxies containing at least 5 star particles and a circular velocity of at least $v_{circ}>60$km/s to account for the overestimate of the low baryonic system numbers \citep{chiu2001, okamoto2008}. 
In order to obtain the ionizing emissivity of the resolved sources in the SPH simulation, we derive the spectral energy distribution (SED) of each source by summing the SEDs of all its star particles. The SED of each star particle has been computed via the population synthesis code \textsc{starburst99} \citep{Leitherer1999} using its mass, stellar metallicity and age. We assume an escape fraction of $f_\mathrm{esc}=0.3$ to ensure that reionization finishes by $z\simeq6$. From two sources at $z\simeq13$ to $\sim55000$ at $z\simeq6$, \textsc{pcrash} computes the evolution of the ionized hydrogen (\HII) and helium fractions (\HeII, \HeIII), as well as the temperature in the IGM, on a $512^3$ grid. The resulting ionization histories of \HII, \HeII\ and \HeIII\ and their ionization maps are shown in Fig. \ref{fig_comparison_HandHe_ionhistory} (black dotted lines), \ref{fig_comparison_HandHe_ionfields}, and \ref{fig_comparison_HandHe_ionfields_HeII}.

\section{Comparing radiative transfer with semi-numerical simulations}
\label{sec_comparison}

\begin{figure}
 \includegraphics[width=0.5\textwidth]{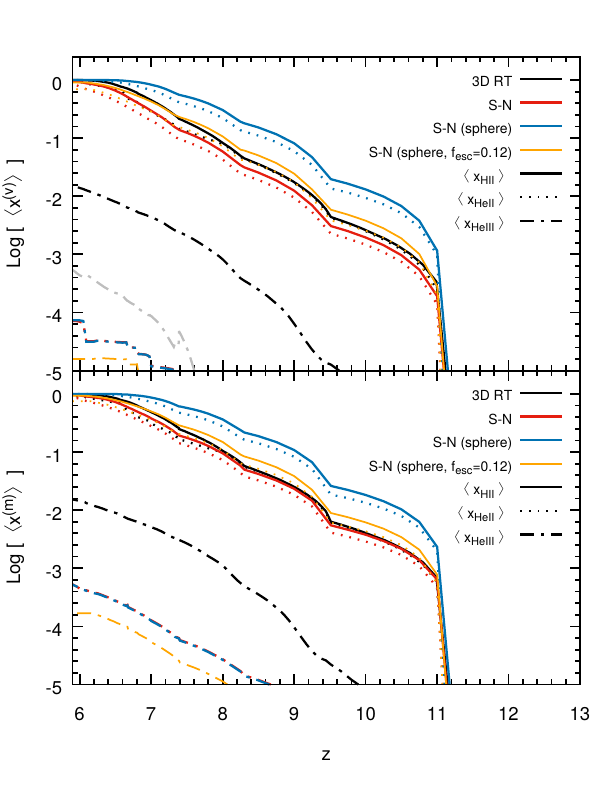}
 \caption{Evolution of the global ionization fractions for \HII\ (solid lines), \HeII\ (dotted lines), and \HeIII\ (dash-dotted lines). Black lines show the results from the RT simulation, while red and blue lines indicate the results from the default and ``ionized sphere'' semi-numerical simulation, all assuming $f_\mathrm{esc}=0.3$. Orange lines depict the results from the ``ionized'' sphere semi-numerical simulation with a reduced ionizing emissivity ($f_\mathrm{esc}=0.12$). The grey line shows the global \HeIII\ fraction from the RT simulation if cells with $\chi_\mathrm{HeIII}<0.99$ are set to zero, i.e. only accounting for fully ionized cells.}
 \label{fig_comparison_HandHe_ionhistory}
\end{figure}

In this Section we describe our set of semi-numerical simulations and compare them to the RT simulation. In particular, we focus on the evolution and distribution of ionized hydrogen (\HII) and singly ionized helium (\HeII) regions.

For this comparison, we post-process the snapshots of the hydrodynamical simulation ($z\simeq13-6$) described in Section \ref{sec_RT} with our semi-numerical reionization simulation code, i.e. assuming the same gas density fields and ionizing sources (location and spectra) for each snapshot. The number of \HI\ ionizing photons of each source is computed according to their spectral energy distribution $L_{\nu}$ as 
\begin{eqnarray}
 \dot{N}_\mathrm{HI,ion}&=& \int_{\nu_\mathrm{HI}}^{\infty} \frac{L_{\nu}}{h\nu}\ \mathrm{d}\nu.
\end{eqnarray}
The number of \HeI\ ionizing photons is derived analogously to Equation \ref{eq_subsec_helium_h_to_he_ratio}
\begin{eqnarray}
 \frac{\dot{N}_\mathrm{HeI,ion}}{\dot{N}_\mathrm{HI,ion}} &=& (1-y) \frac{\alpha_\mathrm{B,HII}}{\alpha_\mathrm{B,HeII}+y \alpha_\mathrm{1,HeII}} \frac{Q_\mathrm{HeI}}{Q_\mathrm{HI}},
\end{eqnarray}
where $Q_\mathrm{HeI} = \int_{\nu_\mathrm{HeI}}^{\infty} L_{\nu}/h\nu\ \mathrm{d}\nu$ and $Q_\mathrm{HI}=\dot{N}_\mathrm{HI,ion}$. The recombination rates $\alpha_\mathrm{B,HII}$, $\alpha_\mathrm{B,HeII}$ and $\alpha_\mathrm{1,HeII}$ are evaluated at the temperature
\begin{eqnarray}
 T &=& 10^5~\mathrm{K}\ \left( \frac{E_\mathrm{HI}/Q_\mathrm{HI}}{41\mathrm{eV}} \right)^{1/4},
\end{eqnarray}
with $E_\mathrm{HI} = \int_{\nu_\mathrm{HI}}^{\infty} L_{\nu}\ \mathrm{d}\nu$, using the relation in \citet{Park_Ricotti2011} (Equation 17).
Similarly, we derive the number of \HeII\ ionizing photons as
\begin{eqnarray}
 \frac{\dot{N}_\mathrm{HeII,ion}}{\dot{N}_\mathrm{HI,ion}} &=& \frac{\alpha_\mathrm{B,HII}}{\alpha_\mathrm{HeIII}} \frac{Q_\mathrm{HeII}}{Q_\mathrm{HI}}
\end{eqnarray}
and $Q_\mathrm{HeII} = \int_{\nu_\mathrm{HeII}}^{\infty} L_{\nu}/h\nu\ \mathrm{d}\nu$.

Using the above numbers of ionizing photons, we run our semi-numerical code from $z\simeq13-6$, following the ionization and recombinations of hydrogen and helium. While our semi-numerical scheme assumes periodic boundaries, the \textsc{pcrash} simulation does not. To imitate the same conditions, we have placed the $512^3$ gas density and ionizing emissivity grids within $1024^3$ grids of mean density and zero values, respectively. This approach ensures that the number of ionization and absorption events at the edges of the box are not increased or decreased by those at the opposite side of the box due to periodic boundary conditions. We note that the maximum smoothing scale does not exceed the original boxsize $80h^{-1}$Mpc (512 cells).

We perform three different runs: our default scheme with an ionizing escape fraction of $f_\mathrm{esc}=0.3$, and two simulation runs with the ``ionized sphere'' flag turned on, with $f_\mathrm{esc}=0.3$ and $0.12$, respectively. The corresponding quantities are shown by the red, blue and orange lines, respectively, in Figures \ref{fig_comparison_HandHe_ionhistory} - \ref{fig_comparison_HandHe_21cm}. We remind the reader that the ``ionized sphere'' flag refers to marking the entire smoothing sphere as ionized instead of the central cell (default scheme) when the ionization criterion in that sphere is met.

\subsection{Ionization history}
\label{sec_comparison_subsec_ionhist}

\begin{figure}
 \includegraphics[width=0.48\textwidth]{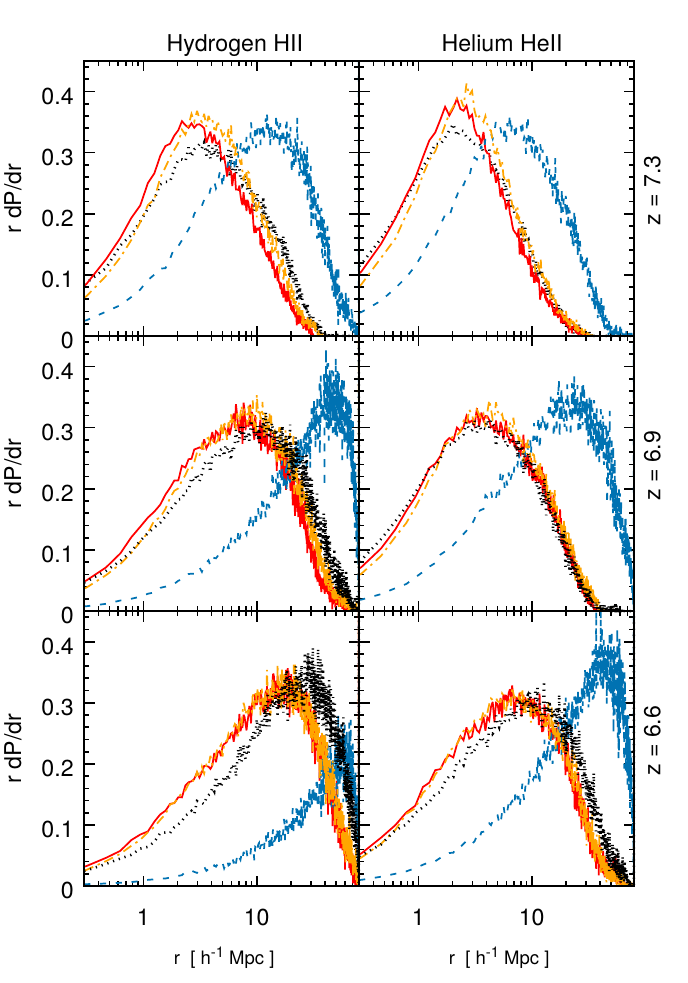}
 \caption{Size distribution of the \HII\ (left column) and \HeII\ (right column) regions at redshifts $7.3$, $6.9$, $6.6$ for the RT simulation (black dotted line), the default (solid red line) and ``ionized'' sphere (dashed blue line) semi-numerical simulation with $f_\mathrm{esc}=0.3$, as well as an ``ionized'' sphere semi-numerical simulation with $f_\mathrm{esc}=0.12$ (dash-dotted orange line). All cells with $\chi_\mathrm{HII}>0.5$ and $\chi_\mathrm{HeII}>0.5$ are considered as ionized.}
 \label{fig_comparison_HandHe_sizedistribution}
\end{figure}

\begin{figure*}
 \includegraphics[width=\textwidth]{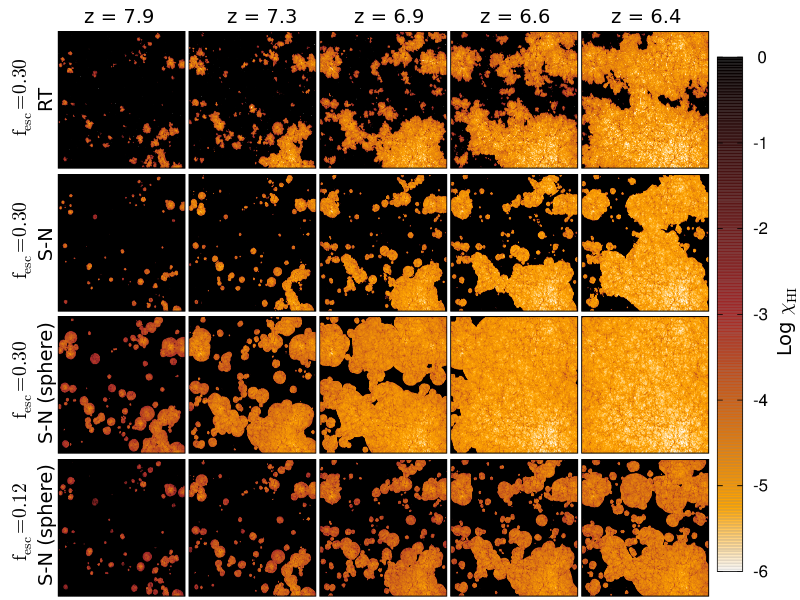}
 \caption{Slices of the \HI\ fields through the $80h^{-1}$Mpc box at redshifts $z=7.9$, $7.3$, $6.9$, $6.6$, $6.4$, corresponding to $\langle \chi_\mathrm{HII}\rangle=0.1$, $0.25$, $0.5$, $0.75$, $0.9$ in the RT simulation, which is shown in the first row. Second and third row show the \HI\ fields of the default and ``ionized'' sphere semi-numerical simulations, both assuming the same value of $f_\mathrm{esc}=0.3$ than the RT simulation. The fourth row depicts the results for an ``ionized'' sphere semi-numerical simulation with a reduced ionizing escape fraction of $f_\mathrm{esc}=0.12$. This $f_\mathrm{esc}$ value has been matched to follow the ionization history of the RT simulation most closely.}
 \label{fig_comparison_HandHe_ionfields}
\end{figure*}

\begin{figure*}
 \includegraphics[width=\textwidth]{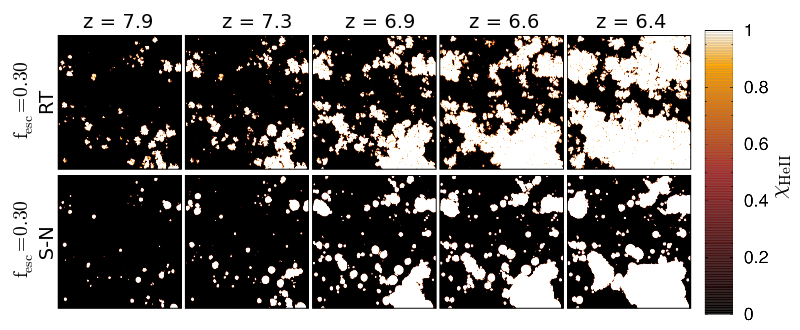}
 \caption{Slices of the \HeII\ fields through the $80h^{-1}$Mpc box at redshifts $z=7.9$, $7.3$, $6.9$, $6.6$, $6.4$, corresponding to $\langle \chi_\mathrm{HeII}\rangle=0.09$, $0.17$, $0.34$, $0.55$, $0.75$ in the RT simulation, which is shown in the first row. The second row depicts the \HeII\ fields of the default semi-numerical simulation, assuming the same value of $f_\mathrm{esc}=0.3$ than the RT simulation.}
 \label{fig_comparison_HandHe_ionfields_HeII}
\end{figure*}

\begin{figure}
 \includegraphics[width=0.48\textwidth]{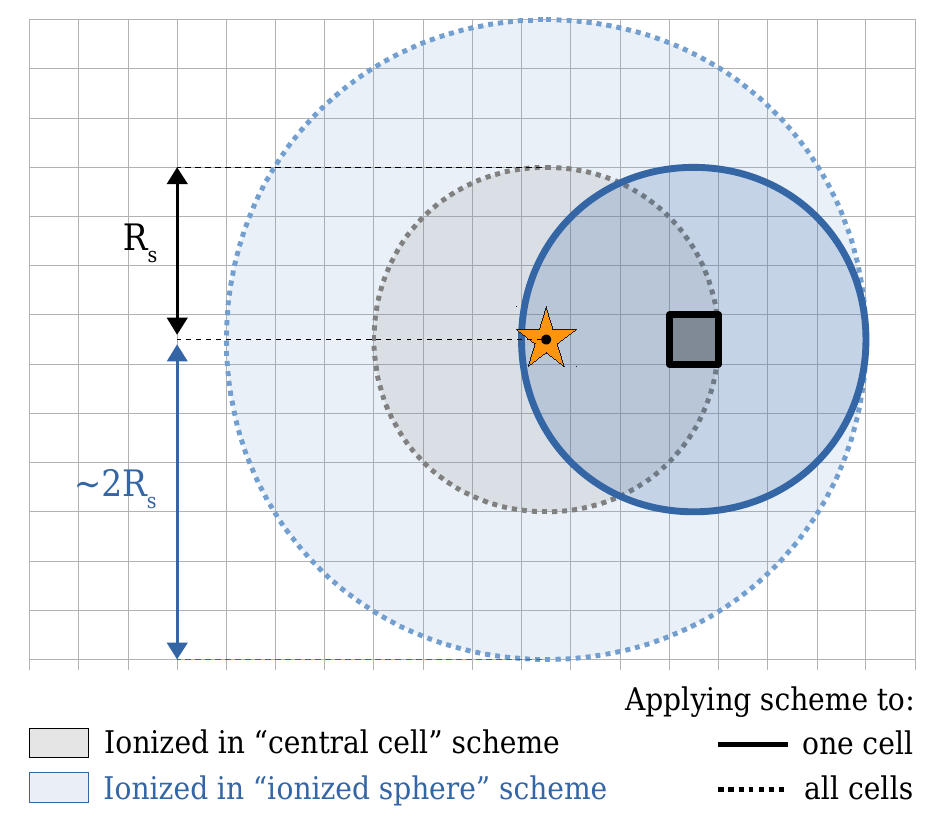}
 \caption{Sketch to illustrate the result of flagging the central cell versus smoothing sphere as ionized when considering a single source. While the ``central cell'' scheme reproduces an ionized regions in agreement with the Str\"omgren sphere, the ``ionized sphere'' scheme ionizes a sphere with twice the Str\"omgren radius.}
 \label{fig_comparison_Honly_sketch}
\end{figure}

We start by comparing the ionization histories and topologies between the different semi-numerical simulations and the RT simulation.

From Fig. \ref{fig_comparison_HandHe_ionhistory} we see that the evolution of the volume (top) and mass (bottom) averaged \HII\ fractions differ among the different semi-numerical approaches. While our default semi-numerical scheme (solid red line) is in reasonable agreement with the results of the RT simulation (solid black line), the ``ionized sphere'' scheme (solid blue line) shifts reionization significantly to higher redshifts by about $\Delta z \simeq0.8$. This difference in the evolution of the ionization fraction can also be seen in Fig. \ref{fig_comparison_HandHe_ionfields}, where we show the \HI\ fields for the RT (first row) and semi-numerical simulations (second, third and fourth row) at redshifts $z=7.9$, $7.3$, $6.9$, $6.6$, $6.4$. In the ``ionized sphere'' scheme (third row) the ionized regions grow considerably faster than in the RT simulation and default scheme; consequently the size distribution of the \HII\ ionized regions (blue line) is shifted significantly towards larger sizes as depicted in the left panels of Fig. \ref{fig_comparison_HandHe_sizedistribution}.

In order to explain the accelerated growth of the ionized regions in the ``ionized sphere'' scheme\footnote{We note that this effect is different to the one discussed in \citet{Zahn2007}. There the authors consider the photon nonconservation of a real space tophat filter in contrast to the photon conservation of a k-space tophat filter, while we focus here on the central cell versus ``ionized sphere'' scheme.}, we consider the simple case of a single source in a homogeneous density field. The default semi-numerical scheme flags the Str\"omgren sphere as ionized (black dashed line in Fig. \ref{fig_comparison_Honly_sketch}), while the ``ionized sphere'' scheme marks a sphere of double the Str\"omgren radius as ionized (blue dashed line in Fig. \ref{fig_comparison_Honly_sketch}). This fact can be understood by considering the cells at the edge of the Str\"omgren sphere that still pass the ionization criterion. Flagging the entire smoothing scale (Str\"omgren radius) around these cells as ionized leads not only to marking the interior of the Str\"omgren sphere as ionized but also the exterior regions (blue solid line in Fig. \ref{fig_comparison_Honly_sketch}). 

An accelerated progression of reionization, as in the ``ionized sphere'' scheme, can be compensated for by a reduced value for the ionizing escape fraction $f_\mathrm{esc}$, and thus the number of ionizing photons. This relation can explain the comparably small $f_\mathrm{esc}$ values of $0.04-0.06$ found in the semi-numerical simulations in \citet{Hassan2016}. Indeed, we find decreasing $f_\mathrm{esc}$ from a value of $0.3$ to $0.12$ in the ``ionized sphere'' scheme yields time evolutions of the volume and mass averaged \HII\ fractions that are in better agreement with the RT simulation (orange solid lines in Fig \ref{fig_comparison_HandHe_ionhistory}). Using our findings as a guide, and assuming the $f_\mathrm{esc}$ values in \citet{Hassan2016} to be about $40$\% of their ``true'' values implies actual $f_\mathrm{esc}$ values in the range of $10-15$\%, being in better agreement with \citet{Robertson2015}, \citet{Bouwens2016} and \citet{Ishigaki2017}. The factor by which the number of ionizing photons needs to be reduced depends on the clustering of the ionizing sources, and thus varies with the mean overdensity of the simulation box. In contrast, for a single source the escape fraction $f_\mathrm{esc}$ would need to be lowered by a factor $1/8$.

Naturally, a reduced ionizing emissivity causes a shallower slope $\mathrm{d}\chi_\mathrm{HI}/\mathrm{d}z$, i.e. a slower progression of reionization, in particular after the ionized regions have started to merge (\avchi$\lsim0.9$). This slower growth is also reflected in the delayed shift of the \HII\ region size distribution in Fig. \ref{fig_comparison_HandHe_sizedistribution}. The difference between the size distribution of the ``ionized sphere'' scheme (orange line) and the RT simulation (black line) increases towards lower redshifts, while the difference between the default scheme (red line) and the RT simulation remains constant on a logarithmic scale. 
The reduced ionizing emissivity also leads to lower photoionization rates and higher residual \HI\ fractions within the ionized regions, as can be seen in Fig. \ref{fig_comparison_HandHe_ionfields}. While the $\chi_\mathrm{HI}$ values in the ionized regions range between $10^{-6}$ to $10^{-4}$ for the original ionizing emissivities (rows $1-3$), they are between $10^{-5}$ to $10^{-3}$ for the reduced $f_\mathrm{esc}$ value scenario (row $4$).

In contrast to the ``ionized sphere'' scheme, our default semi-numerical scheme leads to slopes $\mathrm{d}\chi_\mathrm{HI}/\mathrm{d}z$ and evolutions of the global \HI\ fractions similar to the RT simulation (see Fig. \ref{fig_comparison_HandHe_ionhistory}). More precisely, we find the ionized hydrogen mass to be approximately equal in the default semi-numerical and RT simulations for $\langle\chi_\mathrm{HI}^{(m)}\rangle>0.9$, while the ionization of the IGM is delayed by $\Delta z\simeq0.2$ in the semi-numerical scheme for $\langle\chi_\mathrm{HI}^{(m)}\rangle<0.9$. This delay in ionizing the IGM originates from the photon nonconservation of the semi-numerical scheme \citep[see Appendix in][]{Zahn2007}. As illustrated in Fig. 11 in \citet{Zahn2007} the nonconservation of photons becomes significant when the ionized regions start to merge, and depends on the size distribution of the ionized regions, the luminosity and bias of the sources, and the rate of merging ionized regions. We find the fraction of photons lost reaches its maximum of $\sim20$\% when reionization enters the ``overlap'' phase \citep{Gnedin2000}, and drops back to a few percent when more than $95$\% of its mass is ionized.
We also find the semi-numerical scheme to deviate more from the volume than the mass averaged \HI\ fractions in the RT simulation. This effect can also be seen in the constant offset of the semi-numerical scheme in the size distribution of the ionized regions in Fig. \ref{fig_comparison_HandHe_sizedistribution}, and is expected. 

In the semi-numerical simulations the extent of the ionized region is based on a spherical average of ionization and absorption events, which makes the method less sensitive to small-scale density variations. Low density tunnels are averaged out by higher density filaments and sheets. This is in contrast to RT simulations, where ionizing photons are able to percolate faster through low density tunnels. The averaging effect causes not only a slower progress of reionization, but also smaller and less fractal surfaces between the ionized and neutral regions (see the ionization maps in Fig. \ref{fig_comparison_HandHe_ionfields}), as well as a broader size distribution of the ionized regions with a higher number of small-scale ionized regions (see Fig. \ref{fig_comparison_HandHe_sizedistribution}).
We note that the (default) semi-numerical scheme has been shown to produce some unphysical shapes when ionized regions merge \citep{Zahn2007}, however we have not visually identified this phenomenon in our cosmological ionization fields.

Neglecting recombinations leads to a slightly better agreement with the RT simulation, as we have verified in a run where the number of recombinations was set to zero. The delay in ionization by the recombinations compensates for the photon nonconservation of the semi-numerical method. \citet{Choudhury2009} points out that a density dependent recombination scheme can lead to unphysical tunnelling effects, i.e. the effect of shadowing behind dense regions, also referred to as the butterfly effect \citep{Ciardi2001}, is not produced. Visual inspections of the ionization maps do not reveal such effects in our simulations, which may be due to the lower luminosities of the sources and/or our implementation: In contrast to \citet{Choudhury2009}, who considers self-shielded regions as fully neutral, our schemes derives a residual \HI\ fraction which still classifies the cells as at least partially ionized (see also Section \ref{sec_model_recombinations}). 

\begin{figure*}
 \includegraphics[width=\textwidth]{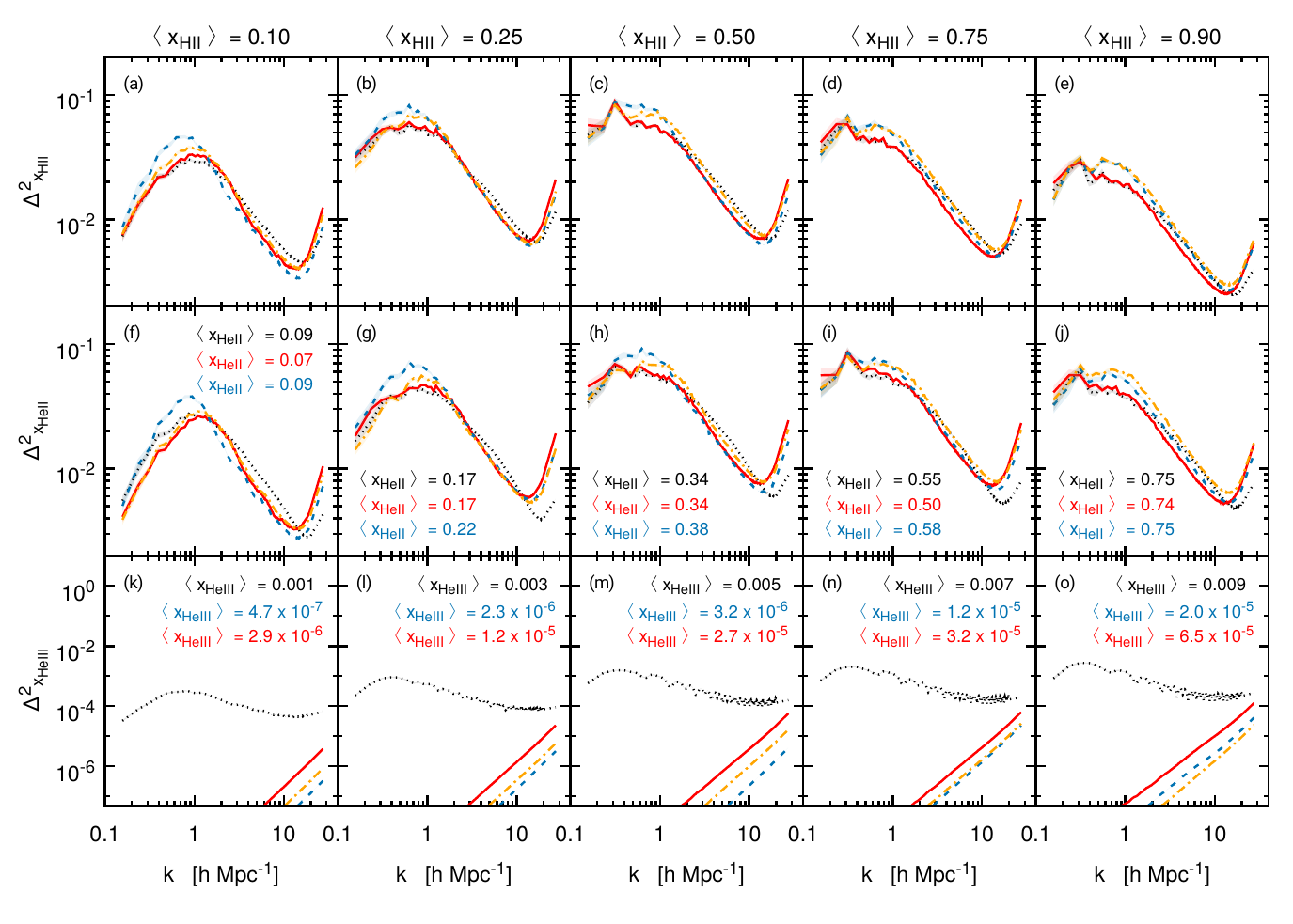}
 \caption{Power spectra of the fluctuations in the \HII\ (first row), \HeII\ (second row), and \HeIII\ fractions at $\langle\chi_\mathrm{HII}\rangle\simeq0.9$, $0.75$, $0.5$, $0.25$, $0.1$ for RT (black dotted) and default (solid red) and ``ionized sphere'' (blue dashed) semi-numerical simulation results. The volume-averaged \HeII\ and \HeIII\ fractions of the different simulations are printed in each panel with their corresponding colour.}
 \label{fig_comparison_HandHe_psion}
\end{figure*}

In our default semi-numerical scheme, we find the volume and mass averaged \HeII\ fractions (red dotted lines) to follow the corresponding \HII\ fractions (solid red lines) with a constant offset. Their time evolutions are very similar to the RT simulation results (black dotted lines), with the mass being in better agreement than the volume averaged \HeII\ fraction.
Despite these similarities, we note that the relation between the average \HeII\ and \HII\ fractions in our semi-numerical scheme differs from the RT simulation in two respects: (1) there is an overall reduced ratio $\langle\chi_\mathrm{HeII}\rangle/\langle\chi_\mathrm{HII}\rangle$, which is probably due to our overestimation of the temperature within the ionized regions. In particular in over-dense regions, where recombination cooling is significant, temperatures are up to an order of magnitude lower in the RT calculations, which causes an increase of $40$\% in the number of \HeII\ ionizing photons. (2) there is no ``drop'' in the ratio $\langle\chi_\mathrm{HeII}\rangle/\langle\chi_\mathrm{HII}\rangle$ at $z\lesssim8$ as in the RT simulation, since the semi-numerical scheme is not as sensitive to a change in spectral shape as the RT and its \HeIII\ fraction is two orders magnitude lower. Both effects, (1) and (2), compensate each other at $z<8$, which leads to better agreement in the size distribution of \HeII\ ionized regions (see right column in Fig. \ref{fig_comparison_HandHe_sizedistribution}).

The discrepancy between the semi-numerical (dashed-dotted red line) and RT (dashed-dotted black line) global \HeIII\ fractions has the following reason: \HeIII\ has an ionization energy of $54.5$eV. For photons with such energies, the \HI\ and \HeI\ ionization cross sections are about an order of magnitude lower than at the \HI\ and \HeI\ ionization thresholds. Thus, \HeII\ ionizing photons have on average a longer mean free path than \HI\ and \HeI\ ionizing photons. Due to these longer mean free paths, we find \HeII\ regions to be partially \HeIII\ ionized and contributing to $\langle \chi_\mathrm{HeIII}\rangle$ in the RT simulation. Indeed, accounting only for cells, that are fully \HeIII\ ionized ($\chi_\mathrm{HeIII}>0.99$) in the RT simulation (gray dashed-dotted line), results in significantly lower $\langle \chi_\mathrm{HeIII}\rangle$ values that are closer to our semi-numerical results. This behaviour is expected, as the semi-numerical scheme only distinguishes between fully neutral, singly or doubly ionized helium. 

We also comment on the evolution of \HeIII\ for the ``ionized sphere'' scheme. Despite flagging the entire sphere when the ionization criterion is met, $\langle \chi_\mathrm{HeIII}\rangle$ shows the same values as our default semi-numerical simulation. This finding reveals that the ionization criterion is only met for smoothing scales of cell size, i.e. only cells containing galaxies become progressively ionized.

\subsection{Size distribution}
\label{sec_comparison_subsec_ion_powerspec}

Due to the complex morphology of the ionized regions, we perform multiple measures for the bubble size distribution. We analyse the distribution of the ionized regions with two statistical methods, the bubble radius distribution\footnote{We compute the bubble radius distribution ($r\ \mathrm{d}P/\mathrm{d}r$) of the ionized regions by randomly selecting $10^5$ ionized cells ($\chi_\mathrm{HII}>0.9$) from where rays are cast in random directions. The rays are propagated until they encounters neutral cells. Their distances are stored as a histogram, from which we derive the bubble radius distributions shown in Fig. \ref{fig_comparison_HandHe_sizedistribution}.} (Fig. \ref{fig_comparison_HandHe_sizedistribution}) and the power spectra of the ionization fields\footnote{We compute the power spectra as $P(k)=\langle\tilde{\Delta}(\vec{k})\ \tilde{\Delta}(-\vec{k})\rangle$, with $\tilde{\Delta}(\vec{k})=1/V \int \mathrm{d}^3x \Delta(\vec{x}) \exp(-i\vec{k}\vec{x})$ being the Fourier transformation of the ionization fields $\Delta(\vec{x})$.} (Fig. \ref{fig_comparison_HandHe_psion}).

Both measures have some features common to all simulations. Firstly, the peak of the bubble radius distribution and the peak of the power spectrum gradually shift from small to larger length scales, which indicates the growth and merger of the ionized regions with cosmic time. Secondly, the amplitude of the power spectrum rises and declines with the variance in the ionization fields, which has its maximum around \avchi$\simeq0.5$ \citep{Lidz2008}. Thirdly, since the cross sections for \HII\ and \HeII\ ionization have similar values at their corresponding ionization energies, \HeII\ follows \HII\ closely for our starburst spectra; thus their power spectra $\Delta^2_{\chi_\mathrm{HII}}$ and $\Delta^2_{\chi_\mathrm{HeII}}$ agree reasonably well with each other.

As noted in the discussion of the ionization histories, we also see from the bubble radius distributions at $z=7.3$, $6.9$, $6.6$ that reionization is slightly delayed in the default semi-numerical scheme and significantly accelerated in the ``ionized sphere'' scheme, when compared to the RT simulation. The power spectra at chosen \avchi\ values also reveal the different size distribution of the ionized regions as it was noted from the ionization maps. The default semi-numerical simulation has less fractal surfaces between ionized and neutral regions than the RT simulation, which leads to less small-scale and more large-scale ionized regions, particularly in the early stages of reionization. This trend is even stronger for the ``ionized sphere'' method, as small-scale fluctuations are even more likely to be erased by neighbouring cells that satisfy the ionization criterion at larger smoothing scales. This result is in agreement with \citet[][comparing RT and hybrid scheme ionization fields]{Zahn2007} and \citet[][comparing RT and \textsc{21cmFAST} 21cm power spectra]{Mesinger2011}. However, it is in contrast to \citet{Zahn2011} and \citet{Majumdar2014}, where MF07 (corresponding to our ``ionized sphere'' with reduced ionizing emissivity'') and Sem-Num (corresponding to our default scheme) show more power on small scales than the RT simulations. A possible reason for this deviation is the time of reionization. While the ionization history in \citet{Zahn2007} with $z_{50\%}=6.9$ is very similar the \avchi\ evolution of our simulation, it is shifted to higher redshifts in \citet{Zahn2011} and \citet{Majumdar2014} ($z_{50\%}=7.6$ and $9.4$, respectively), where density structures are less pronounced.

While the power spectra of the \HII\ and \HeII\ ionization fields are very similar for the RT and semi-numerical simulations, they differ considerably for the \HeIII\ ionization fields.
In the case of the RT simulation, we compare $\Delta^2_{\chi_\mathrm{HeIII}}$ to $\Delta^2_{\chi_\mathrm{HII}}$ and $\Delta^2_{\chi_\mathrm{HeII}}$, and find its amplitude is decreased by two orders of magnitude and its peak shifted to larger scales. The decrease in power arises from the partial \HeIII\ ionization of the \HeII\ ionized regions.  
The relative enhancement of large-scale power, however, is due to the longer mean free paths of \HeIII\ ionizing photons. The longer mean free paths cause a smoothing of the \HeIII\ ionization fractions within the helium ionized regions, so to speak.
However, the shape of the \HeIII\ power spectra of the semi-numerical simulations greatly differ from the RT simulation and the \HII\ and \HeII\ power spectra: the power spectrum drops quickly towards larger scales, which indicates that the \HeIII\ regions are considerably smaller than the \HeII\ regions. As reionization proceeds, $\Delta^2_{\chi_\mathrm{HeIII}}$ increases but retains its shape; this is the result of an increasing number of \HeIII\ ionized regions of similar size. 

We also note that the power of $\Delta^2_{\chi_\mathrm{HeIII}}$ in the ``ionized sphere'' scheme for the original (blue dashed lines) and reduced (orange dash-dotted lines) ionizing emissivity is continuously lower than in our default scheme (red solid lines). In the case of the original ionizing emissivity, this effect is due to the accelerated reionization of the IGM and the fact that we consider times at the same $\langle\chi_\mathrm{HII}\rangle$ values. However, in the case of the reduced ionizing escape fraction, the reduced ionizing emissivity decreases the number of cells meeting the \HeIII\ ionization criteria.

\subsection{Correlations between reionization redshift \& density}
\label{sec_comparison_subsec_crosscorr_zion_dens}

\begin{figure}
 \includegraphics[width=0.48\textwidth]{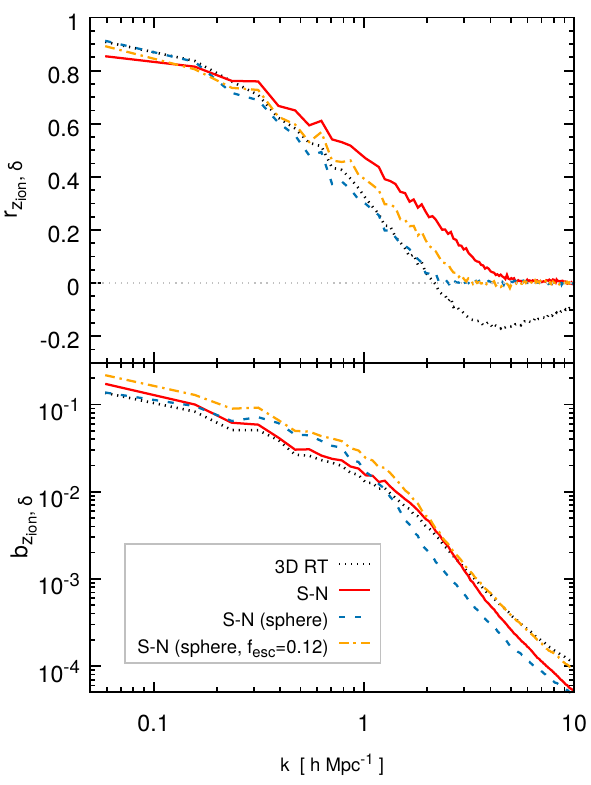}
 \caption{Cross correlation coefficients between the redshifts of reionization and the gas density fields (top) and their corresponding bias (bottom). Black dotted lines show the results for the RT simulation. Red solid and blue dashed lines represent the results for the default and ``ionized'' sphere semi-numerical simulations, respectively, while orange dash-dotted lines depict the ``ionized'' sphere scheme with a reduced $f_\mathrm{esc}$ value of $0.12$.}
 \label{fig_comparison_HandHe_cross_zion_dens}
\end{figure}

In order to investigate the differences between RT and semi-numerical simulations in their ionization topology, we compute the cross correlation coefficient between the gas density and the reionization redshift fields. For this purpose, we construct the reionization redshift field $z_\mathrm{ion}$ by storing the redshift at which it ionizes for each cell, i.e. exceeds a threshold of $\chi_\mathrm{HI}=0.5$.
Next we define the dimensionless fluctuation field of the gas density, $\delta_\rho=\rho(\vec{x})/\bar{\rho}-1$, and the reionization redshift, $\delta_z=z_\mathrm{ion}(\vec{x})/\bar{z}_\mathrm{ion}-1$, with $\bar{\rho}$ being the mean density and $\bar{z}_\mathrm{ion}$ the mean reionization redshift in the simulation box.
Finally, we compute the cross correlation coefficient between the density and reionization redshift fluctuation fields,
\begin{eqnarray}
 r(k) &=& \frac{P_{AB}(k)}{\sqrt{P_{A}(k) P_{B}(k)}},
 \label{eq_sec_comparison_crosscorreff}
\end{eqnarray}
and the linear bias,
\begin{eqnarray}
 b_{AB}(k)&=& \sqrt{\frac{P_{B}(k)}{P_{A}(k)}}.
\end{eqnarray}
$P_{AB}(k) = \langle \tilde{\Delta}_A (\vec{k})\ \tilde{\Delta}_B(-\vec{k}) \rangle$ describes the cross power spectrum between the reionization redshifts and the density fields, and $P_{A}(k)$ and $P_{B}(k)$ are their respective power spectra. The cross correlation coefficient ranges between $-1$ and $1$, and indicates either a correlation ($r(k)>0$), no correlation ($r(k)=0$), or an anti-correlation ($r(k)<0$) between the two datasets.
In Fig. \ref{fig_comparison_HandHe_cross_zion_dens} we show the cross correlation functions $r_{\mathrm{z_{ion}},\rho}$ and the bias factor $b_{\mathrm{z_{ion}},\rho}$ of the RT and semi-numerical simulations in the top and bottom panel, respectively.

In the case of the RT simulation (black dotted lines), the positive cross correlation coefficient $r_{\mathrm{z_{ion}},\rho}$ indicates a correlation between the reionization redshift and the gas density on larger scales ($\lsim 3h^{-1}$Mpc). This correlation indicates a picture wherein over-dense regions hosting the ionizing sources become ionized first and under-dense regions last. However, such behaviour is not true on smaller scales ($\gsim 3h^{-1}$Mpc) where $r_{\mathrm{z_{ion}},\rho}$ becomes negative. This anti-correlation between the time of ionization and gas density indicates an outside-in component in the ionization topology, i.e. low density regions (e.g. voids) are ionized before higher density regions (e.g. filaments). In summary, $r_{\mathrm{z_{ion}},\rho}$ of our RT simulation corresponds to an inside-out-middle ionization topology, where ionization fronts move from dense environments containing the ionizing sources to under-dense voids and finally percolate to the self-shielded filaments. We note that our RT results are in agreement with the \textsc{C$^2$-Ray} results in \citet{Majumdar2014}, with the anti-correlation emerging at the same scale, but have a linear bias $b_{\mathrm{z_{ion}},\rho}$ that is a factor $\sim3$ smaller.

Similar to the RT simulation, we can see from Fig. \ref{fig_comparison_HandHe_cross_zion_dens} that the semi-numerical simulations also exhibit a large-scale correlation which decreases towards smaller scales, i.e. reionization proceeds in an inside-out fashion.
However, as we have already seen in the ionization maps in Fig. \ref{fig_comparison_HandHe_ionfields}, the largest differences between our RT and semi-numerical simulations show up on small scales: in the semi-numerical schemes the times of ionization and the density become uncorrelated and not anti-correlated as in the RT simulation. The reason for this uncorrelation lies in the nature of the ionization criterion. It depends on averaged quantities instead of directional quantities as in the RT. Furthermore, since the semi-numerical simulations miss the outside-in ionization topology component, they show stronger correlation on intermediate scales. Similar results have been reported in \citet{Majumdar2014}, again our default scheme agrees well with their corresponding scheme (Sem-Num).

The differences between the semi-numerical schemes, particularly the drop in correlation towards smaller scales, are due to either the different methods of determining the ionization of a cell, or the ionizing emissivity and the corresponding duration of reionization. In the former case, marking only the central cell (default scheme, red solid line) instead of the entire smoothing sphere (``ionized sphere'' scheme, blue dashed line) results in a stronger correlation between ionization and density, as exceeding the ionization threshold in a cell is more confined to its environment. In the latter case, decreasing the ionizing emissivity (e.g. orange dash-dotted versus blue dashed line) lengthens the duration of reionization and, being more sensitive to ionizing sources that form in the emerging density peaks at later times, and enhances the correlation between reionization redshift and density. This trend is in agreement with the findings in \citet{Battaglia2013} (see their Fig. 2).

\subsection{Correlation between the ionization fields}
\label{sec_comparison_subsec_crosscorr_ion}

\begin{figure}
 \includegraphics[width=0.48\textwidth]{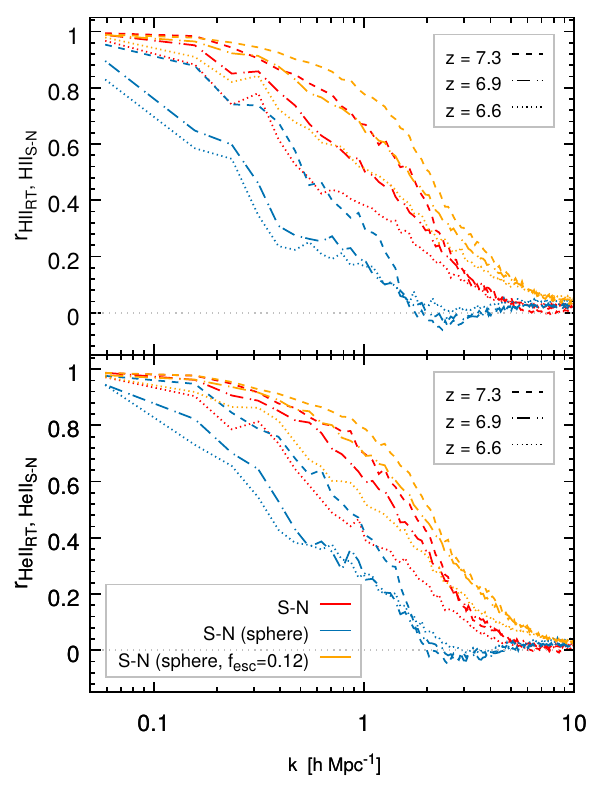}
 \caption{The top (bottom) panel shows the cross correlation coefficients between the \HII\ (\HeII) ionization fields of the RT and the respective semi-numerical simulation at different redshifts. Red and blue lines represent the cross correlations with the default and ``ionized'' sphere semi-numerical simulations, respectively, while orange lines depict the ``ionized'' sphere scheme with a reduced $f_\mathrm{esc}$ value of $0.12$.}
 \label{fig_comparison_HandHe_cross_ion_rt_sn}
\end{figure}

In order to quantify the agreement between the RT and the semi-numerical schemes, we calculate the cross correlation coefficient between the RT and semi-numerical ionization fields according to Equation \ref{eq_sec_comparison_crosscorreff}, with $A$ and $B$ representing the RT and semi-numerically simulated ionization fields.
The results for the cross correlation coefficients of the \HII\ and \HeII\ ionization fields are shown in the top and bottom panel of Fig. \ref{fig_comparison_HandHe_cross_ion_rt_sn}, respectively. 
We find the ionization fields of the RT and semi-numerical simulations to be strongly correlated on large scales ($k\lsim1$ h Mpc$^{-1}$), while the strength of their correlation decreases considerably towards smaller scales, and even becomes uncorrelated in some cases. The drop in the cross correlation coefficients $r_\mathrm{HII_{RT},HII_{S-N}}$ and $r_\mathrm{HeII_{RT},HeII_{S-N}}$ is caused by multiple reasons: Firstly, in the RT simulation, the more fractal shape of the ionization boundaries leads to more small scale structures in the ionization fields (see Fig. \ref{fig_comparison_HandHe_ionfields} and \ref{fig_comparison_HandHe_ionfields_HeII}). Secondly, as reionization proceeds (dashed to dash-dotted to dotted lines), the ionized regions in the RT and semi-numerical schemes merge at different points in space and time, and consequently develop increasingly different shapes with time.

The cross correlation coefficients $r_\mathrm{HII_{RT},HII_{S-N}}$ and $r_\mathrm{HeII_{RT},HeII_{S-N}}$ depend on the agreement of the \avchi\ values of the RT and semi-numerical simulations: the larger the difference in \avchi, the weaker the cross correlation. Plotting $r_\mathrm{HII_{RT},HII_{S-N}}$ and $r_\mathrm{HeII_{RT},HeII_{S-N}}$ at the same \avchi\ values, however, yields similar correlation strengths. 
It is striking that the ``ionized sphere'' scheme with reduced ionizing emissivity (orange lines) shows a stronger correlation than our default scheme (red lines). As can be seen from the ionization maps in Fig. \ref{fig_comparison_HandHe_ionfields}, the former better mimics the bubble shape of the ionized regions in the RT simulation. The ionized regions in the RT simulation deviate only weakly from spheres, as their ionization fronts are predominantly delayed by narrow filamentary density structure.
However, if the source distribution at a chosen \avchi\ value deviates significantly from the source distribution in the RT simulation, the cross correlation strength becomes even weaker (see the ``ionized sphere'' scheme, blue lines). 

\begin{figure*}
 \includegraphics[width=\textwidth]{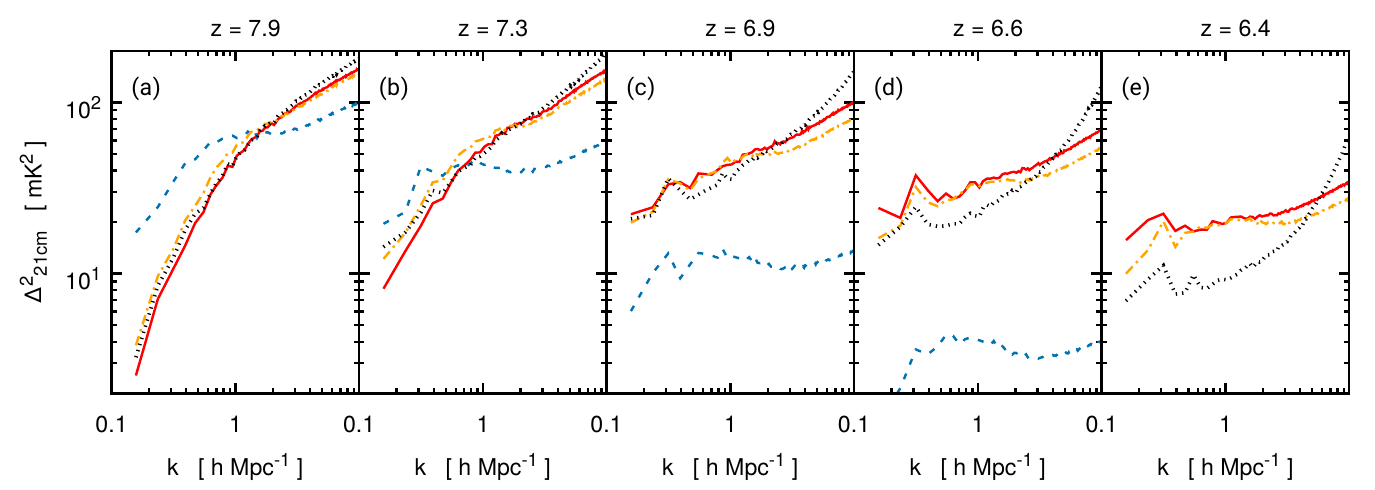}
 \caption{21cm power spectrum at redshifts $z=7.9$, $7.3$, $6.9$, $6.6$, $6.4$ for the RT simulation (black dotted line), the default (solid red line) and ``ionized'' sphere (dashed blue line) semi-numerical simulation with $f_\mathrm{esc}=0.3$, as well as an ``ionized'' sphere semi-numerical simulation with $f_\mathrm{esc}=0.12$ (dash-dotted orange line).}
 \label{fig_comparison_HandHe_21cm}
\end{figure*}

We note that despite using the same ionizing emissivity and gas density fields in the RT and semi-numerical simulations, we find lower $r(k)$ values than \citet{Zahn2007}. We believe that the reason lies in the different ionization topologies of the RT simulations. While the RT simulation in \citet{Zahn2007} proceeds inside-out with over-dense regions being ionized before under-dense regions, our RT simulation exhibits an inside-out-middle behaviour similar to the simulation in \citet{Finlator2009}, where filaments are ionized last due to their increased number of recombinations. The delayed ionization of filaments weakens the cross correlation on scales of the filamentary structure. This relation is supported by the good agreement of our default scheme (at the same \avchi\ values) with the Sem-Num scheme in \citet{Majumdar2014}, with their RT simulation also having a less pronounced inside-out ionization topology.

\subsection{Power spectra of the 21cm signal}
\label{sec_comparison_subsec_21cm_powerspec}

In order to estimate the accuracy of semi-numerical reionization schemes on observables, we compare the 21cm signal power spectra of our semi-numerical and RT simulations. The 21cm differential brightness temperature in each cell of the simulation box is calculated as \citep[e.g.][]{Iliev2012}
\begin{eqnarray}
 \delta T_b (\vec{x}) &=& T_0\ \langle \chi_\mathrm{HI} \rangle \ \left(1+\delta(\vec{x})\right) \ \left(1+\delta_\mathrm{HI}(\vec{x})\right),
 \label{eq_21cm_brightness_temperature}
\end{eqnarray}
with
\begin{eqnarray}
 T_0&=&28.5 \left( \frac{1+z}{10} \right)^{1/2} \frac{\Omega_b}{0.042} \frac{h}{0.073} \left( \frac{\Omega_m}{0.24} \right)^{-1/2} \mathrm{mK}.
 \label{eq_T0}
\end{eqnarray}
$h$ is the Hubble parameter, $\Omega_b$ and $\Omega_m$ the baryonic and matter densities respectively,  $1+\delta(\vec{x})=\rho(\vec{x})/\langle\rho\rangle$ the local gas density in terms of the average gas density, and $1+\delta_\mathrm{HI}(\vec{x})=\chi_\mathrm{HI}(\vec{x})/\langle \chi_\mathrm{HI}\rangle$ the local \HI\ density fraction in terms of the average \HI\ density in the simulation box.

We show the power spectra of the 21cm differential brightness temperature fluctuations at different stages of reionization in Fig. \ref{fig_comparison_HandHe_21cm}. The evolution of the simulated 21cm power spectra is in agreement with previous findings \citep[e.g.][]{Lidz2008}: (1) a drop in power on scales that are larger than the ionized regions, (2) large scale power peaking around \avchi$\simeq0.5$ when the variance of the ionization field is at its maximum, and (3) an overall decrease in power as $\langle \delta T_b \rangle$ drops and reionization completes.

The differences between the RT and semi-numerical simulations in the 21cm power spectra shown in Fig. \ref{fig_comparison_HandHe_21cm} arise from two effects. On the one hand they differ in the distribution of the ionized regions (see the ionization maps in Fig. \ref{fig_comparison_HandHe_ionfields} and bubble radius distribution in Fig. \ref{fig_comparison_HandHe_sizedistribution}), and on the other hand they differ in their \avchi\ values. While semi-numerical simulations show a less fractal structure of the ionization fronts, which reduces the power on small scales, they proceed slower or faster than the RT simulation (black dotted lines) due to their nonconservation of photons. The ``ionized sphere'' scheme with the original ionizing emissivity (blue dashed lines) represents the most extreme case of both these effects and does not agree with our RT results at all.
The default and ``ionized sphere'' schemes with reduced ionizing emissivity reproduce the 21cm power spectra adequately, with a larger deviation in the final stages of reionization, when the differences in their \avchi\ values become larger.

Among the semi-numerical models, the ``ionized sphere'' schemes possesses relatively more power on intermediate scales than our default scheme. This difference is clearly seen when 21cm power spectra are compared at the same \avchi\ values. Furthermore, scales of increased power rise as the ionizing escape fraction is decreased. These effects can be explained with the increasing number of sources that form in density peaks throughout hierarchical structure formation, and their reduced ionizing emissivities that lead to smaller sized ionized regions.

\vspace{0.3cm}

In summary, for \HII\ and \HeII, our default semi-numerical approach reproduces the ionization histories and size distributions of the ionized regions adequately. Though, the photon non-conserving character of the scheme \citep{Zahn2007} causes reionization (\HII\ and \HeII) to be delayed by $\Delta z\sim0.2$. Since the scheme also misses any line of sight dependencies of the ionizing radiation, it shows firstly slightly less (more) small (large) scale structure than our RT simulation, and secondly a stronger inside-out ionization topology.
In contrast, the ``ionized sphere'' semi-numerical scheme leads to a considerably accelerated reionization of the IGM ($\Delta z\sim0.8$). Reducing the ionizing emissivity in this scheme yields a better distribution of the ionization fields than our default scheme, however, results in too high residual \HI\ fractions in the ionized regions and a shallower slope $\mathrm{d}\chi/\mathrm{d}z$ in the ionization history $\chi(z)$ than found in the RT simulation.

Finally, the extent of \HeIII\ ionized regions is only accurate when fully double ionized cells are considered. By construction our method fails to follow the partial ionization of \HeII\ as the RT does. This will be subject to future work.

\section{Discussion and Conclusions}
\label{sec_conclusions}

We present a newly implemented semi-numerical scheme that follows the ionization of hydrogen and helium in the IGM, tracking the ionization fields for \HII, \HeII\ and \HeIII\ for a given set of ionizing sources and gas density field. Similar to the approach in \citet{Furlanetto2004}, our scheme identifies ionized regions by considering the ratio between the number of ionization and absorption events on different spatial smoothing scales. In addition, it includes a description for the number of recombinations for all species, and the spatially dependent \HI\ photoionization rate essential to derive the residual \HI\ fraction in hydrogen ionized regions. We ensure the correct ionization of helium in the IGM by taking secondary ionization of hydrogen from recombining helium into account when the number of \HeI\ ionizing photons are computed. The accuracy of the number of \HeI\ ionizing photons is further improved by considering the recombination rates of each source at the IGM temperature that is predicted by its spectral energy distribution \citep[see relation in][]{Park_Ricotti2011}.

We compare our new semi-numerical scheme to a RT simulation run with \textsc{pcrash} \citep{Partl2011,Hutter2014}. In contrast to previous works \citep[e.g.][]{Zahn2007,Zahn2011,Majumdar2014}, we do not adjust the ionizing emissivities in the semi-numerical simulations to reproduce the ionization history ($\langle\chi\rangle(z)$) of the RT run, but assume the same source locations and ionizing emissivities. This allows us to test the accuracy of these semi-numerical reionization schemes across the full duration of reionization, in particular their uncertainty in constraining parameters such as the escape fraction of ionizing photons $f_\mathrm{esc}$ and ionizing emissivity of galaxies.

When comparing the number of ionization and absorption events in a region and the ionization criterion is met, our semi-numerical code has the option of choosing between flagging only the central cell (default) or the entire smoothing sphere (``ionized sphere'') as ionized. We resimulate the RT simulation with our semi-numerical code and follow the progress of reionization for both options. 

Turning on the ``ionized sphere'' flag results in an accelerated reionization scenario despite assuming the same production rate of ionizing photons and escape fraction $f_\mathrm{esc}$. For our reionization scenario, which is in agreement with recent Planck observations \citep{Planck2016} for the RT simulation, we find the ``ionized sphere'' scheme to shorten the duration of reionization by $\Delta z\simeq0.8$. In other words, to obtain the same reionization history as in the RT simulation, we need to reduce the escape fraction $f_\mathrm{esc}$ to $40$\% of its initial value. Fitting the $f_\mathrm{esc}$ value for the RT ionization history yields ionization fields that are in agreement with those of the RT simulation, but increases the residual \HI\ fractions in the ionized regions by an order of magnitude, and alters the slope $\mathrm{d}\chi/\mathrm{d}z$ of the redshift evolution $\chi(z)$.

For the ``central cell'' (or default) scheme, we find the \HII\ and \HeII\ ionization histories and size distributions of the \HII\ and \HeII\ ionized regions to be in reasonable agreement with the RT simulations. The small delay ($\Delta z\sim 0.2$) in completing reionization is due to the photon nonconservation of the real space tophat filter \citep{Zahn2007}. The averaging of the density and ionizing emissivity fields, which increases (decreases) the effective density in under-dense (over-dense) regions, results in more connected ionized regions, suppressing (enhancing) the small (large) scale variations in the ionization fields, and strengthens the inside-out character of the ionization topology. Thus, the power spectra of the 21cm signal at selected redshifts are flatter and have increasingly more power in the final stages of reionization.

Our findings regarding the redshift evolution of reionization highlight that the ``central cell'' method yields more accurate results than the ``ionized sphere'' scheme. This is particularly important when parameters that impact the duration of reionization are constrained, such as the escape fraction of ionizing photons, are constrained. 

Furthermore, we presented the first semi-numerical scheme that follows hydrogen and helium reionization simultaneously. While our prescription for the number of \HeI\ ionizing photons reproduces the \HeII\ ionized regions in the RT simulation, the extent of the \HeIII\ regions is only accurate when fully double ionized cells are considered. The latter is expected, as by construction our scheme can not follow the partial ionization of \HeII\ as RT simulations do. This limitation will be subject to further work.

It is important to note some caveats, however.
Firstly, our description of recombinations does not account for unresolved density inhomogeneities. While this should not alter the reionization scenario strongly, as recombination time scales of the IGM are of the order of the Hubble time at these redshifts, the number of recombinations in dense regions, e.g. Lyman Limit systems, may be underestimated and their ionization delayed. 
Secondly, we assume a constant escape fraction of ionizing photons across all galaxies. This is probably an unlikely scenario, because an increasing number of simulations have pointed to a halo mass dependent value \citep{Ferrara2013, Kimm2014, Kimm2017}.
However, this does not affect the main points of this paper: we have shown that our newly developed semi-numerical reionization scheme yields comparable results to RT simulations for \HII\ and \HeII. Furthermore, an ``ionized sphere'' scheme should not be employed when parameters regarding the duration of reionization are to be constrained.

Our new code computes the ionization fields from the gas density grids and a list of sources containing their ionizing emissivity. This makes it highly versatile, as the modelling of the ionizing sources (i.e. luminosity and spectral energy distribution) is left to the user. Thus it will be straight forward to couple this code to semi-analytic galaxy formation models or hydrodynamical simulations. The modular fashion of the code also allows the user to replace the current descriptions of recombinations and the photoionization rate. Coupled to a semi-analytic galaxy formation model, this versatile implementation of the reionization of the IGM will allow us to investigate the diversity of reionization topologies and their signatures in the power spectrum of the \HI\ 21cm signal, and the cross correlations between the 21cm signal and high-redshift galaxies.

\section*{Acknowledgements} 
The author thanks the referee for their insightful and constructive comments that improved the quality of the paper. Furthermore, the author thanks Darren Croton, Manodeep Sinha, Simon Mutch and Jacob Seiler for helpful discussions and comments, and Robert Dzudzar for proof reading. 
AH is supported under the Australian Research Council's Discovery Project funding scheme (project number DP150102987). Parts of this research were conducted by the Australian Research Council Centre of Excellence for All Sky Astrophysics in 3 Dimensions (ASTRO 3D), through project number CE170100013.
 
\appendix

\section{Normalisation of the photoionization rate} 
\label{sec_app2}
We define the photoionization rates in the mean free path based model as
\begin{eqnarray}
 \Gamma_{V/\lambda}(r) &=& \sigma_0 \frac{\alpha}{\alpha+\beta} \frac{\dot{N}_\mathrm{ion,V/\lambda}}{V}\ \lambda_\mathrm{mfp},
\end{eqnarray}
whilst in our flux based model as
\begin{eqnarray}
 \Gamma_A (r) &=& \sigma_0 \frac{\alpha}{\alpha+\beta} \frac{\dot{N}_\mathrm{ion,A}}{A(r)}.
\end{eqnarray}
In order to find a normalisation for the expression in the mean free path based model, we consider the photoionization rate within a sphere of radius $\lambda_\mathrm{mfp}$. Then the photoionization rate is given by
\begin{eqnarray}
 \frac{4\pi R^3}{3} \ \Gamma_{V/\lambda}(R)&=& \sigma_0\ \frac{\alpha}{\alpha+\beta}\ \dot{N}_\mathrm{ion,V/\lambda}\ \frac{\lambda_\mathrm{mfp}}{V}\ \frac{4\pi R^3}{3},
\end{eqnarray}
while for the second we find
\begin{eqnarray}
 &4\pi &\int_0^R \mathrm{d}r\ r^2\ \Gamma_A(r) \nonumber\\
 &=& \int_0^R \mathrm{d}r\ r^2\ \sigma_0\ \frac{\alpha}{\alpha+\beta}\ \dot{N}_\mathrm{ion,A}\ \frac{e^{-r/\lambda_\mathrm{mfp}}}{k r^2} \nonumber\\
 &=& \sigma_0\ \frac{\alpha}{\alpha+\beta}\ \dot{N}_\mathrm{ion,A}\ \frac{1}{k}\ \lambda_\mathrm{mfp}\ \left(1-e^{-R/\lambda_\mathrm{mfp}}\right),
\end{eqnarray}
with $k$ being a normalisation constant.
We caution the reader that $\dot{N}_\mathrm{ion,V/\lambda}$ and $\dot{N}_\mathrm{ion,A}$ are not equal.
$\dot{N}_\mathrm{ion,A}$ is the ionizing rate in one cell, while $\dot{N}_\mathrm{ion,V/\lambda}$ corresponds to the ionizing rate $\dot{N}_\mathrm{ion,A}$ distributed over $N$ cells. The number of cells is at most the number of cells in a sphere with radius $\lambda_\mathrm{mfp}$, however, it is mostly smaller. While $N=4\pi/3 \lambda_\mathrm{mfp}^3$ is true for ionized cells close to ionization fronts, it is not applicable to cells deeper inside the ionized regions and $N$ is smaller. To account for this effect, we introduce a free parameter, $f<1$, and 
\begin{eqnarray}
 \dot{N}_\mathrm{ion,V/\lambda} &=& \frac{\dot{N}_\mathrm{ion,A}}{N}\ =\ \dot{N}_\mathrm{ion,A}\ \frac{V}{\frac{4\pi}{3}(f \lambda_\mathrm{mfp})^3}.
\end{eqnarray}
This relation between $\dot{N}_\mathrm{ion,V/\lambda}$ and $\dot{N}_\mathrm{ion,A}$, and the assumption that the photoionization rate within a sphere of radius $\lambda_\mathrm{mfp}$ should be equal for both approaches, leads to
\begin{eqnarray}
 k &=&  \frac{\dot{N}_\mathrm{ion,A}}{\dot{N}_\mathrm{ion,V/\lambda}}\ \frac{V}{\frac{4\pi}{3} R^3}\ \left(1-e^{-R/\lambda_\mathrm{mfp}}\right) \nonumber \\
 &=& f^3 \left(\frac{\lambda_\mathrm{mfp}}{R}\right)^3\ \left(1-e^{-R/\lambda_\mathrm{mfp}}\right) \\
 &\stackrel{R=\lambda_\mathrm{mfp}}{=}& f^3 \left(1-e^{-1}\right).
\end{eqnarray}
We have chosen $f=0.7$.

\section{Hydrogen only reionization simulation} 
\label{sec_app_honly_reionsim}

\subsection{Radiative Transfer} 
\label{sec_app_honly_RT}

\begin{figure*}
 \includegraphics[width=0.9\textwidth]{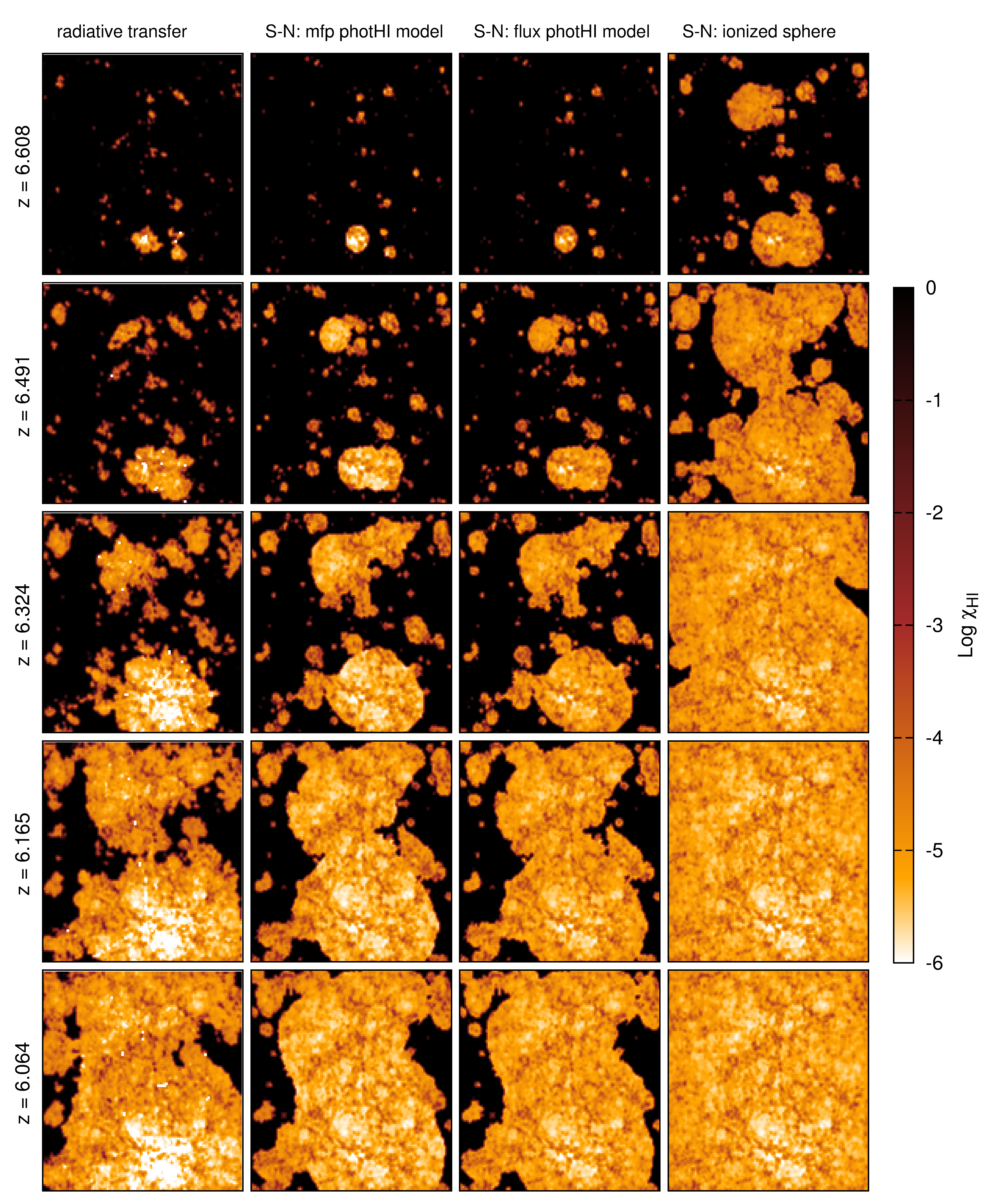}
 \caption{Slices of the \HI\ fields through the centre of the $80h^{-1}$Mpc box at different redshifts $z=6.61$, $6.49$, $6.32$, $6.17$, $6.06$ (rows). The first column shows the \HI\ fields from the RT simulations with \textsc{pcrash}, whereas \avchi$\simeq0.9$, $0.75$, $0.5$, $0.25$, $0.1$ from top to bottom. The second and third column show the \HI\ fields  of our default semi-numerical code for photoionization model 1 ($\Gamma\propto\lambda_\mathrm{mfp}$) and 2 ($\Gamma\propto e^{r/\langle\lambda_\mathrm{mfp}\rangle}/r^2$), respectively, with \avchi$\simeq0.95$, $0.81$, $0.58$, $0.36$, $0.25$ from top to bottom. The location and sizes of the ionized regions (yellow/white) agree for the RT and default semi-numerical simulations. The fourth column shows the \HI\ fields of the ``ionized sphere'' option of our semi-numerical code assuming photoionization model 1, \avchi$\simeq0.77$, $0.30$, $0.05$, $0.005$, $0.001$ from top to bottom.}
 \label{fig_comparison_Honly_HIfields}
\end{figure*}

We evaluated the accuracy of our semi-numerical schemes with respect to another RT hydrogen-only simulation. This simulation is described in \citet{Hutter2014} and follows the time and spatial evolution of hydrogen (no helium). The $z\simeq6.7$ snapshot of the hydrodynamical simulation described in Section \ref{sec_RT} is post-processed with \textsc{pcrash}. However, in this simulation, we only consider sources with at least $10$ star particles and $M_h>10^{9.2}\msun$. Their ionizing emissivity is derived from the the spectral energy distribution (SED) of each source by summing the SEDs of all its star particles. The SED of each star particle has been computed via the population synthesis code \textsc{starburst99} \citep{Leitherer1999} using its mass, stellar metallicity and age. In this comparison we focus on the simulation with an escape fraction of ionizing photons from the galaxy of $f_\mathrm{esc}=0.5$. 
Starting from a completely neutral IGM, \textsc{pcrash}  has been run until the IGM global hydrogen fraction drops below \avchi$\simeq10^{-4}$, computing the evolution of the ionization field for the ionizing radiation of 31855 ``resolved'' galaxies on a $128^3$ grid. We store snapshots at different \avchi\ values ($0.9$, $0.75$, $0.5$, $0.25$, $0.1$) throughout reionization. The respective ionization fields are shown in the left column of Fig. \ref{fig_comparison_Honly_HIfields}.

\subsection{Comparison}
\label{sec_app_honly_comp}

\begin{figure*}
 \includegraphics[width=1.0\textwidth]{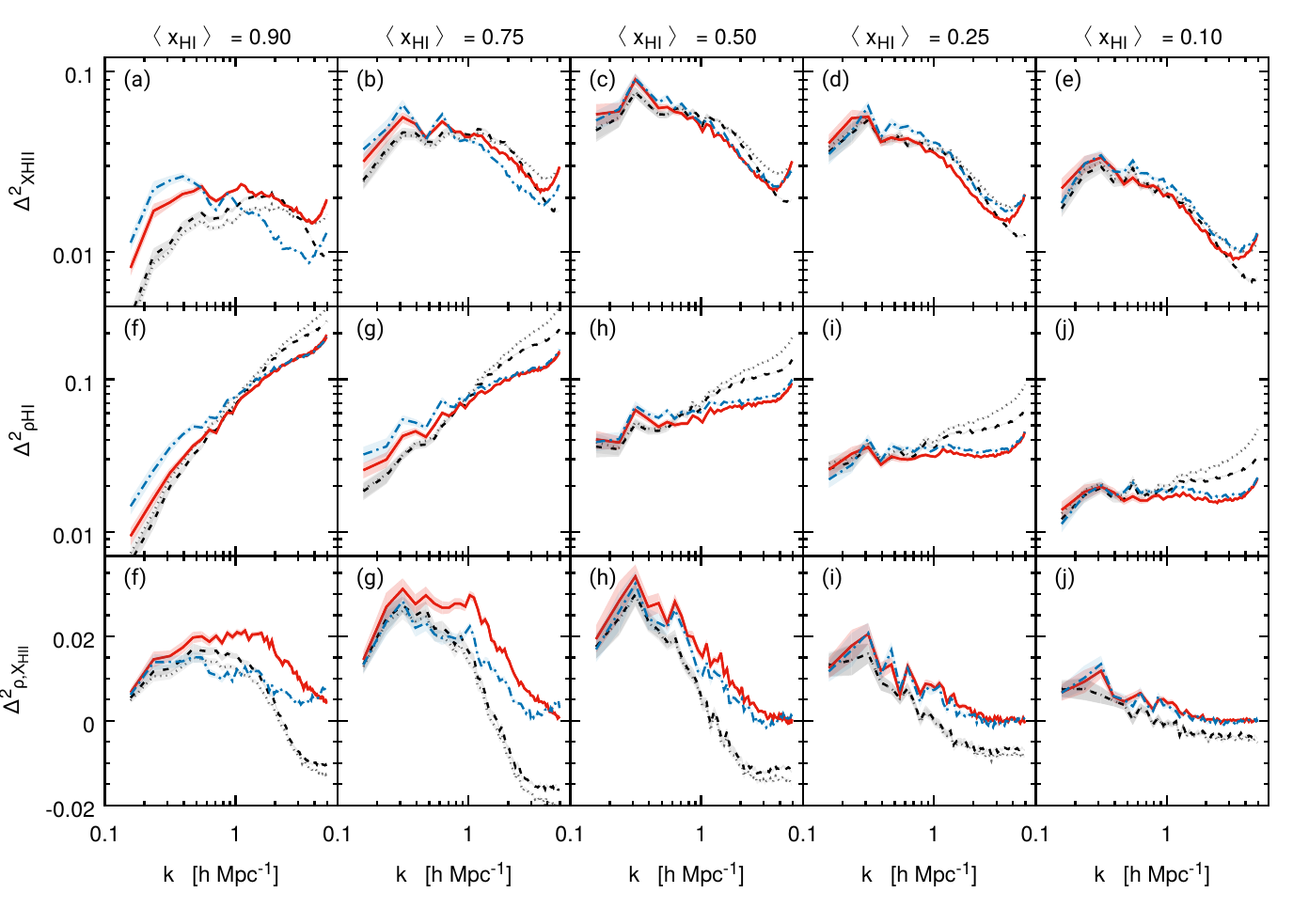}
 \caption{Power spectra of the ionization fields (first row), neutral density fluctuations (second row), and cross power spectrum of the ionization and density fields at \avchi$\simeq0.9$, $0.75$, $0.5$, $0.25$, $0.1$ for RT (black dashed), default (red solid) and ``ionized sphere'' (blue dash-dotted) semi-numerical simulation results. The black dotted lines represent the results when the ionization fraction in all cells in the RT simulation with $\chi_\mathrm{HII}<0.5$ are set to zero. The shaded transparent regions represent the statistical variances of the k-space bins.}
 \label{fig_comparison_Honly_ps}
\end{figure*}

\subsubsection{Ionization history \& morphology}
\label{sec_app_honly_comp_ion_morphology}

We perform the reionization simulation described in Section \ref{sec_app_honly_RT} with the different versions of our semi-numerical code. Similar to Section \ref{sec_comparison}, we use the same gas density fields and ionizing sources; the ionizing photons $\dot{N}_\mathrm{ion}$ for each source are computed according to Equation \ref{eq_subsec_esf_nion}. Again, RT density and ionizing emissivity $128^3$ grids are placed within a $256^3$ grid with mean density and zero values, respectively, to account for the non-periodic boundary conditions of the RT simulation. 
The different versions of our semi-numerical code are: (1) central cell method with the mean free path approach based photoionization model (see Section \ref{sec_model_par_photHI_model1}), (2) central cell method with the flux approach based photoionization model (see Section \ref{sec_model_par_photHI_model2}), and (3) the ``ionized sphere'' model with the flux approached photoionization model. The respective ionization fields are shown in Fig. \ref{fig_comparison_Honly_HIfields} at redshifts $z=6.61$, $6.49$, $6.32$, $6.17$, $6.06$, which correspond to \avchi $\simeq0.9$, $0.75$, $0.5$, $0.25$, $0.1$ in the RT simulation, with the first, second, tird, and fourth column displaying the results for the RT simulation, the semi-numerical simulation for the mean free path and flux based photoionization models, the ``ionized sphere'' semi-numerical scheme, respectively.

Comparing the redshift evolution of reionization as well as the morphology of the ionized region, our results agree well with our findings in Section \ref{sec_comparison}: The ``ionized sphere'' scheme (column 4) results in a significantly faster ionization of the IGM, while the ``central cell'' method (column 2 and 3) yields more comparable results to the RT simulation (column 1). The different size distributions of the ionized regions are also reflected in the power spectra of the ionization fields (first row in Fig. \ref{fig_comparison_Honly_ps}) and the neutral density fluctuations (second row in Fig. \ref{fig_comparison_Honly_ps}), with the semi-numerical schemes, in particular the ``ionized sphere'' scheme, lacking small-scale real-space power. We remark that the thin grey dotted line represents results from the RT simulation when all cells with $\chi_\mathrm{HII}>0.5$ are set to $1$, while all cells with $\chi_\mathrm{HII}\leq0.5$ are set to $0$. This underlines that the reduced small-scale power in the power spectra of the ionization fields and ionized density fluctuations in the RT simulations are due to a shallower decrease of the ionization fraction towards the ionization boundaries. Enhanced small-scale power of semi-numerical schemes with sharp ionization fronts have been also found in \citet{Zahn2007,Zahn2011} and \citet{Majumdar2014} for their CPS scheme.

The cross correlation power spectra between the ionization and density fields (third row in Fig. \ref{fig_comparison_Honly_ps}) underline and reproduce our findings in Section \ref{sec_comparison_subsec_crosscorr_zion_dens}. While the RT simulation shows a correlation between the ionization fraction and the gas density on large scales, it becomes (1) anti-correlated at increasing scales as reionization proceeds, with $\sim 3h^{-1}$Mpc at \avchi$=0.9$ to $\sim6h^{-1}$Mpc at \avchi$=0.1$, and (2) varies in its anti-correlation strength peaking around \avchi$\sim0.3-0.5$. As the ionized regions grow, larger self-shielded filaments can be traced. During the ``overlap'' phase \citep{Gnedin2000} the butterfly shape of the ionized regions \citep{Ciardi2001,Choudhury2009}, which contributes to the anti-correlation, becomes less pronounced. In addition, similar to our findings in Section \ref{sec_comparison_subsec_crosscorr_zion_dens}, the semi-numerical schemes show a stronger correlation between ionization and density field, and arises from the ionization criterion being dependent on the smoothed density field instead of the density and the direction-dependent radiation field in each cell.

\subsubsection{Photoionization models}
\label{sec_app_honly_comp_photionmodels}

Finally, we shortly discuss the results of our two photoionization models implemented in our semi-numerical scheme. Both only affect the \HI\ fraction within but not the extent of the ionized regions. While the mean free path based approach derives the photoionization rate from the mean free path in the cell given by the maximum filtering size at which the cell becomes ionized, the flux based approach assumes an exponential decay of the ionizing flux with the average mean free path. As we can see from the first three columns in Fig. \ref{fig_comparison_Honly_HIfields}, both models approximately recover the residual \HI\ distribution within the ionized regions. 
The \HI\ residual fraction in the mean free path based approach (second column in Fig. \ref{fig_comparison_Honly_HIfields}) is very sensitive to the filtering scale at which cells are identified as ionized. This fact leads to large areas of higher photoionization rate values (lower \HI\ fractions) and small areas of comparably lower photoionization rate values (higher \HI\ fractions), particularly visible at \avchi$>0.5$. In addition, this method can result in low \HI\ values close to the ionization fronts, which differs qualitatively from the RT simulation results.
This drawback is addressed in the flux based approach. With the photoionization rate being dependent on the distance of the ionizing sources, the increase in the \HI\ fraction further away from the ionizing sources and closer to the ionization fronts is ensured. However, due to the photoionization rate decaying exponentially with the {\it average} mean free path, the spatial variance in the mean free path is neglected and as such we do not find as large differences in the \HI\ residual fractions as in the RT simulation. Nevertheless, the flux based approach qualitatively reproduces the residual \HI\ fraction distribution within the ionized regions.

\bibliographystyle{mn2e}
\bibliography{gridmodel_rev1}

\begin{thebibliography}{77}
\expandafter\ifx\csname natexlab\endcsname\relax\def\natexlab#1{#1}\fi

\bibitem[{{Ba{\~n}ados} {et~al}\mbox{.}(2016){Ba{\~n}ados}, {Venemans},
  {Decarli}, {Farina}, {Mazzucchelli}, {Walter}, {Fan}, {Stern}, {Schlafly},
  {Chambers}, {Rix}, {Jiang}, {McGreer}, {Simcoe}, {Wang}, {Yang}, {Morganson},
  {De Rosa}, {Greiner}, {Balokovi{\'c}}, {Burgett}, {Cooper}, {Draper},
  {Flewelling}, {Hodapp}, {Jun}, {Kaiser}, {Kudritzki}, {Magnier}, {Metcalfe},
  {Miller}, {Schindler}, {Tonry}, {Wainscoat}, {Waters}, \&
  {Yang}}]{Banados2016}
{Ba{\~n}ados} E. {et~al.}, 2016, \apjs, 227, 11

\bibitem[{{Battaglia} {et~al}\mbox{.}(2013){Battaglia}, {Trac}, {Cen}, \&
  {Loeb}}]{Battaglia2013}
{Battaglia} N., {Trac} H., {Cen} R., {Loeb} A., 2013, \apj, 776, 81

\bibitem[{{Bauer} {et~al}\mbox{.}(2015){Bauer}, {Springel}, {Vogelsberger},
  {Genel}, {Torrey}, {Sijacki}, {Nelson}, \& {Hernquist}}]{Bauer2015}
{Bauer} A., {Springel} V., {Vogelsberger} M., {Genel} S., {Torrey} P.,
  {Sijacki} D., {Nelson} D., {Hernquist} L., 2015, \mnras, 453, 3593

\bibitem[{{Bolton} \& {Haehnelt}(2007)}]{Bolton_Haehnelt2007}
{Bolton} J.~S., {Haehnelt} M.~G., 2007, \mnras, 382, 325

\bibitem[{{Bouwens} {et~al}\mbox{.}(2017){Bouwens}, {Oesch}, {Illingworth},
  {Ellis}, \& {Stefanon}}]{Bouwens2017}
{Bouwens} R.~J., {Oesch} P.~A., {Illingworth} G.~D., {Ellis} R.~S., {Stefanon}
  M., 2017, \apj, 843, 129

\bibitem[{{Bouwens} {et~al}\mbox{.}(2016){Bouwens}, {Smit}, {Labb{\'e}},
  {Franx}, {Caruana}, {Oesch}, {Stefanon}, \& {Rasappu}}]{Bouwens2016}
{Bouwens} R.~J., {Smit} R., {Labb{\'e}} I., {Franx} M., {Caruana} J., {Oesch}
  P., {Stefanon} M., {Rasappu} N., 2016, \apj, 831, 176

\bibitem[{{Chiu} {et~al}\mbox{.}(2001){Chiu}, {Gnedin}, \&
  {Ostriker}}]{chiu2001}
{Chiu} W.~A., {Gnedin} N.~Y., {Ostriker} J.~P., 2001, \apj, 563, 21

\bibitem[{{Choudhury} \& {Ferrara}(2007)}]{Choudhury_Ferrara2007}
{Choudhury} T.~R., {Ferrara} A., 2007, \mnras, 380, L6

\bibitem[{{Choudhury} {et~al}\mbox{.}(2009){Choudhury}, {Haehnelt}, \&
  {Regan}}]{Choudhury2009}
{Choudhury} T.~R., {Haehnelt} M.~G., {Regan} J., 2009, \mnras, 394, 960

\bibitem[{{Ciardi} {et~al}\mbox{.}(2012){Ciardi}, {Bolton}, {Maselli}, \&
  {Graziani}}]{Ciardi2012}
{Ciardi} B., {Bolton} J.~S., {Maselli} A., {Graziani} L., 2012, \mnras, 423,
  558

\bibitem[{{Ciardi} {et~al}\mbox{.}(2001){Ciardi}, {Ferrara}, {Marri}, \&
  {Raimondo}}]{Ciardi2001}
{Ciardi} B., {Ferrara} A., {Marri} S., {Raimondo} G., 2001, \mnras, 324, 381

\bibitem[{{D'Aloisio} {et~al}\mbox{.}(2017){D'Aloisio}, {Upton Sanderbeck},
  {McQuinn}, {Trac}, \& {Shapiro}}]{Daloisio2017}
{D'Aloisio} A., {Upton Sanderbeck} P.~R., {McQuinn} M., {Trac} H., {Shapiro}
  P.~R., 2017, \mnras, 468, 4691

\bibitem[{{Dixon} {et~al}\mbox{.}(2014){Dixon}, {Furlanetto}, \&
  {Mesinger}}]{Dixon2014}
{Dixon} K.~L., {Furlanetto} S.~R., {Mesinger} A., 2014, \mnras, 440, 987

\bibitem[{{Fan} {et~al}\mbox{.}(2006){Fan}, {Strauss}, {Becker}, {White},
  {Gunn}, {Knapp}, {Richards}, {Schneider}, {Brinkmann}, \&
  {Fukugita}}]{Fan2006}
{Fan} X. {et~al.}, 2006, \aj, 132, 117

\bibitem[{{Ferrara} \& {Loeb}(2013)}]{Ferrara2013}
{Ferrara} A., {Loeb} A., 2013, \mnras, 431, 2826

\bibitem[{{Finlator} {et~al}\mbox{.}(2016){Finlator}, {Oppenheimer},
  {Dav{\'e}}, {Zackrisson}, {Thompson}, \& {Huang}}]{Finlator2016}
{Finlator} K., {Oppenheimer} B.~D., {Dav{\'e}} R., {Zackrisson} E., {Thompson}
  R., {Huang} S., 2016, \mnras, 459, 2299

\bibitem[{{Finlator} {et~al}\mbox{.}(2009){Finlator}, {{\"O}zel}, {Dav{\'e}},
  \& {Oppenheimer}}]{Finlator2009}
{Finlator} K., {{\"O}zel} F., {Dav{\'e}} R., {Oppenheimer} B.~D., 2009, \mnras,
  400, 1049

\bibitem[{{Furlanetto} {et~al}\mbox{.}(2004){Furlanetto}, {Zaldarriaga}, \&
  {Hernquist}}]{Furlanetto2004}
{Furlanetto} S.~R., {Zaldarriaga} M., {Hernquist} L., 2004, \apj, 613, 1

\bibitem[{{Ghara} {et~al}\mbox{.}(2015{\natexlab{a}}){Ghara}, {Choudhury}, \&
  {Datta}}]{Ghara2015a}
{Ghara} R., {Choudhury} T.~R., {Datta} K.~K., 2015{\natexlab{a}}, \mnras, 447,
  1806

\bibitem[{{Ghara} {et~al}\mbox{.}(2015{\natexlab{b}}){Ghara}, {Datta}, \&
  {Choudhury}}]{Ghara2015b}
{Ghara} R., {Datta} K.~K., {Choudhury} T.~R., 2015{\natexlab{b}}, \mnras, 453,
  3143

\bibitem[{{Giallongo} {et~al}\mbox{.}(2015){Giallongo}, {Grazian}, {Fiore},
  {Fontana}, {Pentericci}, {Vanzella}, {Dickinson}, {Kocevski}, {Castellano},
  {Cristiani}, {Ferguson}, {Finkelstein}, {Grogin}, {Hathi}, {Koekemoer},
  {Newman}, \& {Salvato}}]{Giallongo2015}
{Giallongo} E. {et~al.}, 2015, \aap, 578, A83

\bibitem[{{Gnedin}(2000)}]{Gnedin2000}
{Gnedin} N.~Y., 2000, \apj, 535, 530

\bibitem[{{Gnedin}(2014)}]{Gnedin2014}
{Gnedin} N.~Y., 2014, \apj, 793, 29

\bibitem[{{Hassan} {et~al}\mbox{.}(2016){Hassan}, {Dav{\'e}}, {Finlator}, \&
  {Santos}}]{Hassan2016}
{Hassan} S., {Dav{\'e}} R., {Finlator} K., {Santos} M.~G., 2016, \mnras, 457,
  1550

\bibitem[{{Hassan} {et~al}\mbox{.}(2017){Hassan}, {Dav{\'e}}, {Mitra},
  {Finlator}, {Ciardi}, \& {Santos}}]{Hassan2017}
{Hassan} S., {Dav{\'e}} R., {Mitra} S., {Finlator} K., {Ciardi} B., {Santos}
  M.~G., 2017, ArXiv e-prints

\bibitem[{{Hutter}(2015)}]{Hutter2015}
{Hutter} A., 2015, PhD thesis

\bibitem[{{Hutter} {et~al}\mbox{.}(2017){Hutter}, {Dayal}, {M{\"u}ller}, \&
  {Trott}}]{Hutter2017}
{Hutter} A., {Dayal} P., {M{\"u}ller} V., {Trott} C.~M., 2017, \apj, 836, 176

\bibitem[{{Hutter} {et~al}\mbox{.}(2014){Hutter}, {Dayal}, {Partl}, \&
  {M{\"u}ller}}]{Hutter2014}
{Hutter} A., {Dayal} P., {Partl} A.~M., {M{\"u}ller} V., 2014, \mnras, 441,
  2861

\bibitem[{{Iliev} {et~al}\mbox{.}(2014){Iliev}, {Mellema}, {Ahn}, {Shapiro},
  {Mao}, \& {Pen}}]{Iliev2014}
{Iliev} I.~T., {Mellema} G., {Ahn} K., {Shapiro} P.~R., {Mao} Y., {Pen} U.-L.,
  2014, \mnras, 439, 725

\bibitem[{{Iliev} {et~al}\mbox{.}(2012){Iliev}, {Mellema}, {Shapiro}, {Pen},
  {Mao}, {Koda}, \& {Ahn}}]{Iliev2012}
{Iliev} I.~T., {Mellema} G., {Shapiro} P.~R., {Pen} U.-L., {Mao} Y., {Koda} J.,
  {Ahn} K., 2012, \mnras, 423, 2222

\bibitem[{{Ishigaki} {et~al}\mbox{.}(2017){Ishigaki}, {Kawamata}, {Ouchi},
  {Oguri}, \& {Shimasaku}}]{Ishigaki2017}
{Ishigaki} M., {Kawamata} R., {Ouchi} M., {Oguri} M., {Shimasaku} K., 2017,
  ArXiv e-prints

\bibitem[{{Kakiichi} {et~al}\mbox{.}(2016){Kakiichi}, {Dijkstra}, {Ciardi}, \&
  {Graziani}}]{Kakiichi2016}
{Kakiichi} K., {Dijkstra} M., {Ciardi} B., {Graziani} L., 2016, \mnras, 463,
  4019

\bibitem[{{Kim} {et~al}\mbox{.}(2013){Kim}, {Wyithe}, {Raskutti}, {Lacey}, \&
  {Helly}}]{Kim2013}
{Kim} H.-S., {Wyithe} J.~S.~B., {Raskutti} S., {Lacey} C.~G., {Helly} J.~C.,
  2013, \mnras, 428, 2467

\bibitem[{{Kimm} \& {Cen}(2014)}]{Kimm2014}
{Kimm} T., {Cen} R., 2014, \apj, 788, 121

\bibitem[{{Kimm} {et~al}\mbox{.}(2017){Kimm}, {Katz}, {Haehnelt}, {Rosdahl},
  {Devriendt}, \& {Slyz}}]{Kimm2017}
{Kimm} T., {Katz} H., {Haehnelt} M., {Rosdahl} J., {Devriendt} J., {Slyz} A.,
  2017, \mnras, 466, 4826

\bibitem[{{Knollmann} \& {Knebe}(2009)}]{Knollmann2009}
{Knollmann} S.~R., {Knebe} A., 2009, \apjs, 182, 608

\bibitem[{{Kulkarni} {et~al}\mbox{.}(2017){Kulkarni}, {Choudhury}, {Puchwein},
  \& {Haehnelt}}]{Kulkarni2017}
{Kulkarni} G., {Choudhury} T.~R., {Puchwein} E., {Haehnelt} M.~G., 2017,
  \mnras, 469, 4283

\bibitem[{{La Plante} \& {Trac}(2016)}]{LaPlante2016}
{La Plante} P., {Trac} H., 2016, \apj, 828, 90

\bibitem[{{La Plante} {et~al}\mbox{.}(2017){La Plante}, {Trac}, {Croft}, \&
  {Cen}}]{LaPlante2017}
{La Plante} P., {Trac} H., {Croft} R., {Cen} R., 2017, \apj, 841, 87

\bibitem[{{Leitherer} {et~al}\mbox{.}(1999){Leitherer}, {Schaerer}, {Goldader},
  {Delgado}, {Robert}, {Kune}, {de Mello}, {Devost}, \&
  {Heckman}}]{Leitherer1999}
{Leitherer} C. {et~al.}, 1999, \apjs, 123, 3

\bibitem[{{Lidz} {et~al}\mbox{.}(2008){Lidz}, {Zahn}, {McQuinn}, {Zaldarriaga},
  \& {Hernquist}}]{Lidz2008}
{Lidz} A., {Zahn} O., {McQuinn} M., {Zaldarriaga} M., {Hernquist} L., 2008,
  \apj, 680, 962

\bibitem[{{Majumdar} {et~al}\mbox{.}(2014){Majumdar}, {Mellema}, {Datta},
  {Jensen}, {Choudhury}, {Bharadwaj}, \& {Friedrich}}]{Majumdar2014}
{Majumdar} S., {Mellema} G., {Datta} K.~K., {Jensen} H., {Choudhury} T.~R.,
  {Bharadwaj} S., {Friedrich} M.~M., 2014, \mnras, 443, 2843

\bibitem[{{Maselli} {et~al}\mbox{.}(2009){Maselli}, {Ciardi}, \&
  {Kanekar}}]{Maselli2009}
{Maselli} A., {Ciardi} B., {Kanekar} A., 2009, \mnras, 393, 171

\bibitem[{{Maselli} {et~al}\mbox{.}(2003){Maselli}, {Ferrara}, \&
  {Ciardi}}]{Maselli2003}
{Maselli} A., {Ferrara} A., {Ciardi} B., 2003, \mnras, 345, 379

\bibitem[{{McQuinn}(2012)}]{McQuinn2012}
{McQuinn} M., 2012, \mnras, 426, 1349

\bibitem[{{McQuinn} {et~al}\mbox{.}(2007){McQuinn}, {Lidz}, {Zahn}, {Dutta},
  {Hernquist}, \& {Zaldarriaga}}]{McQuinn2007}
{McQuinn} M., {Lidz} A., {Zahn} O., {Dutta} S., {Hernquist} L., {Zaldarriaga}
  M., 2007, \mnras, 377, 1043

\bibitem[{{McQuinn} \& {Switzer}(2009)}]{McQuinn2009}
{McQuinn} M., {Switzer} E.~R., 2009, \prd, 80, 063010

\bibitem[{{Meerburg} {et~al}\mbox{.}(2013){Meerburg}, {Dvorkin}, \&
  {Spergel}}]{Meerburg2013}
{Meerburg} P.~D., {Dvorkin} C., {Spergel} D.~N., 2013, \apj, 779, 124

\bibitem[{{Mesinger} {et~al}\mbox{.}(2015){Mesinger}, {Aykutalp}, {Vanzella},
  {Pentericci}, {Ferrara}, \& {Dijkstra}}]{Mesinger2015}
{Mesinger} A., {Aykutalp} A., {Vanzella} E., {Pentericci} L., {Ferrara} A.,
  {Dijkstra} M., 2015, \mnras, 446, 566

\bibitem[{{Mesinger} \& {Furlanetto}(2007)}]{Mesinger2007}
{Mesinger} A., {Furlanetto} S., 2007, \apj, 669, 663

\bibitem[{{Mesinger} {et~al}\mbox{.}(2011){Mesinger}, {Furlanetto}, \&
  {Cen}}]{Mesinger2011}
{Mesinger} A., {Furlanetto} S., {Cen} R., 2011, \mnras, 411, 955

\bibitem[{{Miralda-Escud{\'e}} {et~al}\mbox{.}(2000){Miralda-Escud{\'e}},
  {Haehnelt}, \& {Rees}}]{Miralda2000}
{Miralda-Escud{\'e}} J., {Haehnelt} M., {Rees} M.~J., 2000, \apj, 530, 1

\bibitem[{{Mutch} {et~al}\mbox{.}(2016){Mutch}, {Geil}, {Poole}, {Angel},
  {Duffy}, {Mesinger}, \& {Wyithe}}]{Mutch2016}
{Mutch} S.~J., {Geil} P.~M., {Poole} G.~B., {Angel} P.~W., {Duffy} A.~R.,
  {Mesinger} A., {Wyithe} J.~S.~B., 2016, \mnras, 462, 250

\bibitem[{{Ocvirk} {et~al}\mbox{.}(2016){Ocvirk}, {Gillet}, {Shapiro},
  {Aubert}, {Iliev}, {Teyssier}, {Yepes}, {Choi}, {Sullivan}, {Knebe},
  {Gottl{\"o}ber}, {D'Aloisio}, {Park}, {Hoffman}, \& {Stranex}}]{Ocvirk2016}
{Ocvirk} P. {et~al.}, 2016, \mnras, 463, 1462

\bibitem[{{Okamoto} {et~al}\mbox{.}(2008){Okamoto}, {Gao}, \&
  {Theuns}}]{okamoto2008}
{Okamoto} T., {Gao} L., {Theuns} T., 2008, \mnras, 390, 920

\bibitem[{{O'Shea} {et~al}\mbox{.}(2015){O'Shea}, {Wise}, {Xu}, \&
  {Norman}}]{Oshea2015}
{O'Shea} B.~W., {Wise} J.~H., {Xu} H., {Norman} M.~L., 2015, \apjl, 807, L12

\bibitem[{{Ouchi} {et~al}\mbox{.}(2017){Ouchi}, {Harikane}, {Shibuya},
  {Shimasaku}, {Taniguchi}, {Konno}, {Kobayashi}, {Kajisawa}, {Nagao}, {Ono},
  {Inoue}, {Umemura}, {Mori}, {Hasegawa}, {Higuchi}, {Komiyama}, {Matsuda},
  {Nakajima}, {Saito}, \& {Wang}}]{Ouchi2017}
{Ouchi} M. {et~al.}, 2017, ArXiv e-prints

\bibitem[{{Paardekooper} {et~al}\mbox{.}(2015){Paardekooper}, {Khochfar}, \&
  {Dalla Vecchia}}]{Paardekooper2015}
{Paardekooper} J.-P., {Khochfar} S., {Dalla Vecchia} C., 2015, \mnras, 451,
  2544

\bibitem[{{Park} \& {Ricotti}(2011)}]{Park_Ricotti2011}
{Park} K., {Ricotti} M., 2011, \apj, 739, 2

\bibitem[{{Parsa} {et~al}\mbox{.}(2017){Parsa}, {Dunlop}, \&
  {McLure}}]{Parsa2017}
{Parsa} S., {Dunlop} J.~S., {McLure} R.~J., 2017, ArXiv e-prints

\bibitem[{{Partl} {et~al}\mbox{.}(2011){Partl}, {Maselli}, {Ciardi}, {Ferrara},
  \& {M{\"u}ller}}]{Partl2011}
{Partl} A.~M., {Maselli} A., {Ciardi} B., {Ferrara} A., {M{\"u}ller} V., 2011,
  \mnras, 414, 428

\bibitem[{{Pawlik} {et~al}\mbox{.}(2017){Pawlik}, {Rahmati}, {Schaye}, {Jeon},
  \& {Dalla Vecchia}}]{Pawlik2017}
{Pawlik} A.~H., {Rahmati} A., {Schaye} J., {Jeon} M., {Dalla Vecchia} C., 2017,
  \mnras, 466, 960

\bibitem[{{Planck Collaboration} {et~al}\mbox{.}(2016){Planck Collaboration},
  {Adam}, {Aghanim}, {Ashdown}, {Aumont}, {Baccigalupi}, {Ballardini},
  {Banday}, {Barreiro}, {Bartolo}, {Basak}, {Battye}, {Benabed}, {Bernard},
  {Bersanelli}, {Bielewicz}, {Bock}, {Bonaldi}, {Bonavera}, {Bond}, {Borrill},
  {Bouchet}, {Boulanger}, {Bucher}, {Burigana}, {Calabrese}, {Cardoso},
  {Carron}, {Chiang}, {Colombo}, {Combet}, {Comis}, {Couchot}, {Coulais},
  {Crill}, {Curto}, {Cuttaia}, {Davis}, {de Bernardis}, {de Rosa}, {de Zotti},
  {Delabrouille}, {Di Valentino}, {Dickinson}, {Diego}, {Dor{\'e}}, {Douspis},
  {Ducout}, {Dupac}, {Elsner}, {En{\ss}lin}, {Eriksen}, {Falgarone}, {Fantaye},
  {Finelli}, {Forastieri}, {Frailis}, {Fraisse}, {Franceschi}, {Frolov},
  {Galeotta}, {Galli}, {Ganga}, {G{\'e}nova-Santos}, {Gerbino}, {Ghosh},
  {Gonz{\'a}lez-Nuevo}, {G{\'o}rski}, {Gruppuso}, {Gudmundsson}, {Hansen},
  {Helou}, {Henrot-Versill{\'e}}, {Herranz}, {Hivon}, {Huang}, {Ili{\'c}},
  {Jaffe}, {Jones}, {Keih{\"a}nen}, {Keskitalo}, {Kisner}, {Knox},
  {Krachmalnicoff}, {Kunz}, {Kurki-Suonio}, {Lagache}, {L{\"a}hteenm{\"a}ki},
  {Lamarre}, {Langer}, {Lasenby}, {Lattanzi}, {Lawrence}, {Le Jeune},
  {Levrier}, {Lewis}, {Liguori}, {Lilje}, {L{\'o}pez-Caniego}, {Ma},
  {Mac{\'{\i}}as-P{\'e}rez}, {Maggio}, {Mangilli}, {Maris}, {Martin},
  {Mart{\'{\i}}nez-Gonz{\'a}lez}, {Matarrese}, {Mauri}, {McEwen}, {Meinhold},
  {Melchiorri}, {Mennella}, {Migliaccio}, {Miville-Desch{\^e}nes}, {Molinari},
  {Moneti}, {Montier}, {Morgante}, {Moss}, {Naselsky}, {Natoli}, {Oxborrow},
  {Pagano}, {Paoletti}, {Partridge}, {Patanchon}, {Patrizii}, {Perdereau},
  {Perotto}, {Pettorino}, {Piacentini}, {Plaszczynski}, {Polastri}, {Polenta},
  {Puget}, {Rachen}, {Racine}, {Reinecke}, {Remazeilles}, {Renzi}, {Rocha},
  {Rossetti}, {Roudier}, {Rubi{\~n}o-Mart{\'{\i}}n}, {Ruiz-Granados},
  {Salvati}, {Sandri}, {Savelainen}, {Scott}, {Sirri}, {Sunyaev}, {Suur-Uski},
  {Tauber}, {Tenti}, {Toffolatti}, {Tomasi}, {Tristram}, {Trombetti},
  {Valiviita}, {Van Tent}, {Vielva}, {Villa}, {Vittorio}, {Wandelt}, {Wehus},
  {White}, {Zacchei}, \& {Zonca}}]{Planck2016}
{Planck Collaboration} {et~al.}, 2016, \aap, 596, A108

\bibitem[{{Qin} {et~al}\mbox{.}(2017){Qin}, {Mutch}, {Poole}, {Liu}, {Duffy},
  {Geil}, {Angel}, {Mesinger}, \& {Wyithe}}]{Qin2017}
{Qin} Y. {et~al.}, 2017, ArXiv e-prints

\bibitem[{{Rahmati} {et~al}\mbox{.}(2013){Rahmati}, {Pawlik}, {Rai{\v c}evic},
  \& {Schaye}}]{Rahmati2013}
{Rahmati} A., {Pawlik} A.~H., {Rai{\v c}evic} M., {Schaye} J., 2013, \mnras,
  430, 2427

\bibitem[{{Robertson} {et~al}\mbox{.}(2015){Robertson}, {Ellis}, {Furlanetto},
  \& {Dunlop}}]{Robertson2015}
{Robertson} B.~E., {Ellis} R.~S., {Furlanetto} S.~R., {Dunlop} J.~S., 2015,
  \apjl, 802, L19

\bibitem[{{Santos} {et~al}\mbox{.}(2010){Santos}, {Ferramacho}, {Silva},
  {Amblard}, \& {Cooray}}]{Santos2010}
{Santos} M.~G., {Ferramacho} L., {Silva} M.~B., {Amblard} A., {Cooray} A.,
  2010, \mnras, 406, 2421

\bibitem[{{Schaye}(2001)}]{Schaye2001}
{Schaye} J., 2001, \apjl, 562, L95

\bibitem[{{Sobacchi} \& {Mesinger}(2014)}]{Sobacchi2014}
{Sobacchi} E., {Mesinger} A., 2014, \mnras, 440, 1662

\bibitem[{{Springel}(2005)}]{Springel2005}
{Springel} V., 2005, \mnras, 364, 1105

\bibitem[{{Takeuchi} {et~al}\mbox{.}(2014){Takeuchi}, {Zaroubi}, \&
  {Sugiyama}}]{Takeuchi2014}
{Takeuchi} Y., {Zaroubi} S., {Sugiyama} N., 2014, \mnras, 444, 2236

\bibitem[{{Thomas} {et~al}\mbox{.}(2009){Thomas}, {Zaroubi}, {Ciardi},
  {Pawlik}, {Labropoulos}, {Jeli{\'c}}, {Bernardi}, {Brentjens}, {de Bruyn},
  {Harker}, {Koopmans}, {Mellema}, {Pandey}, {Schaye}, \&
  {Yatawatta}}]{Thomas2009}
{Thomas} R.~M. {et~al.}, 2009, \mnras, 393, 32

\bibitem[{{Trac} \& {Cen}(2007)}]{Trac2007}
{Trac} H., {Cen} R., 2007, \apj, 671, 1

\bibitem[{{Weigel} {et~al}\mbox{.}(2015){Weigel}, {Schawinski}, {Treister},
  {Urry}, {Koss}, \& {Trakhtenbrot}}]{Weigel2015}
{Weigel} A.~K., {Schawinski} K., {Treister} E., {Urry} C.~M., {Koss} M.,
  {Trakhtenbrot} B., 2015, \mnras, 448, 3167

\bibitem[{{Wyithe} \& {Loeb}(2003)}]{Wyithe_Loeb2003}
{Wyithe} J.~S.~B., {Loeb} A., 2003, \apjl, 588, L69

\bibitem[{{Zahn} {et~al}\mbox{.}(2007){Zahn}, {Lidz}, {McQuinn}, {Dutta},
  {Hernquist}, {Zaldarriaga}, \& {Furlanetto}}]{Zahn2007}
{Zahn} O., {Lidz} A., {McQuinn} M., {Dutta} S., {Hernquist} L., {Zaldarriaga}
  M., {Furlanetto} S.~R., 2007, \apj, 654, 12

\bibitem[{{Zahn} {et~al}\mbox{.}(2011){Zahn}, {Mesinger}, {McQuinn}, {Trac},
  {Cen}, \& {Hernquist}}]{Zahn2011}
{Zahn} O., {Mesinger} A., {McQuinn} M., {Trac} H., {Cen} R., {Hernquist} L.~E.,
  2011, \mnras, 414, 727

\end{thebibliography}

\appendix

\end{document}